\documentclass[10pt,a4paper]{article}
\usepackage[utf8]{inputenc}
\usepackage{amsmath}
\usepackage{amsfonts}
\usepackage{hyperref}
\usepackage{amssymb}
\usepackage{makeidx}
\usepackage{cite}
\usepackage{bm}
\usepackage{geometry}
\usepackage{cite}

\usepackage{doc}

\usepackage{url}

\usepackage{graphicx}
\usepackage{epstopdf}
\begin{document}
\begin{titlepage}

~~\\

\vspace*{0cm}
    \begin{Large}
    \begin{bf}
       \begin{center}
         {Comments on spinning OPE blocks in AdS$_{3}$/CFT$_{2}$}
       \end{center}
    \end{bf}   
    \end{Large}
    
  \vspace{0.7cm}
\begin{center}   
Suchetan Das,\footnote
            {
e-mail address : 
suchetan1993@gmail.com}

\vspace{0.3cm}
 {\it Ramakrishna Mission Vivekananda Educational and Research Institute, Belur Math, Howrah-711202, West Bengal, India}
\end{center}

\vspace{0.7cm}
\begin{abstract}
We extend the work of \cite{Ferrara:1971vh},\cite{Ferrara:1973vz}, to obtain an integral expression of OPE blocks for spinning primaries in CFT$_{2}$. We observe, when the OPE blocks are made out of conserved spinning primaries, the integral becomes a product of two copies of weighted AdS$_{2}$ fields, smeared along geodesics. In this way, conserved current OPE blocks in CFT$_{2}$ have a different representation in terms of AdS$_{2}$ geodesic operators, in stead of viewing them as AdS$_{3}$ geodesic operators. We also show, how this representation can be related to AdS$_{3}$ massless higher spin fields through HKLL bulk field reconstruction. Using this picture, we consistently obtain the closed form expression of four point spinning conformal block as a product of two AdS$_{2}$ Geodesic Witten diagrams.
\end{abstract}
 \end{titlepage}

\pagenumbering{arabic}

\tableofcontents

\section{Introduction} 

Recently, using embedding space formalism \cite{Dirac:1936fq}-\cite{Penedones:2016voo}, a generalization of scalar Geodesic Witten diagram(GWD) \cite{Hijano:2015zsa} to spinning geodesic Witten diagram has been proposed \cite{Dyer:2017zef} - \cite{Costa:2018mcg}\footnote{In the first order Hilbert-Palatini formalism of gravity, the holographic dual conformal blocks are also studied using open Wilson networks \cite{Bhatta:2016hpz}-\cite{Bhatta:2018gjb}.}. This computes conformal partial waves(CPW)(\cite{Mack:1969rr}-\cite{Ferrara:1973yt}) with spinning exchange operator as well as spinning external operators in CFT$_{d}$ (\cite{Costa:2011mg}-\cite{Karateev:2017jgd}). For scalar GWD we have:
\begin{align}
W_{\Delta.0}(x_{1},x_{2},x_{3},x_{4}) \propto \int_{\gamma_{12}}d\lambda\int_{\gamma_{34}} d\lambda'G_{b\partial}(y(\lambda),x_{1}) G_{b\partial}(y(\lambda),x_{2})G_{bb}(y(\lambda),y(\lambda');\Delta)G_{b\partial}(y(\lambda'),x_{3})G_{b\partial}(y(\lambda'),x_{4})
\end{align}
where $G_{b\partial}$, the bulk to boundary propagator and $G_{bb}$, the bulk to bulk propagator in AdS$_{d+1}$ are integrated over two geodesics $\gamma_{12}$ and $\gamma_{34}$. This prescription could be understood from a more fundamental identification between the so called `OPE blocks' $B_{k}^{ij}(x_{1},x_{2})$(building block of an OPE) in the CFT and `Geodesic operators' in the bulk \cite{Czech:2016xec}\cite{deBoer:2016pqk},\cite{daCunha:2016crm}.\footnote{See \cite{Karch:2017fuh}-\cite{Cresswell:2018mpj} for dual description of OPE blocks in states other than vacuum.}

In CFT$_{2}$ \cite{Dolan:2003hv}-\cite{Osborn:2012vt}, the expression for spinning conformal block involves a product of two CFT$_{1}$ conformal blocks $k_{h}(z)$ \cite{Dolan:2011dv}.
\begin{align}\label{2dclosed}
W_{h,\bar{h}}^{h_{i},\bar{h}_{i}}(z,\bar{z}) = k_{h}(z)k_{\bar{h}}(\bar{z})+k_{\bar{h}}(z)k_{h}(\bar{z}), \\
z=\frac{z_{12}z_{34}}{z_{13}z_{24}}; \quad k_{h}(z) = z^{h} {}_2F_1\left(h-h_{12},h+h_{34};2h;z\right)
\end{align}
This (anti)holomorphic factorization implies that in 2D CFT, in stead of using AdS$_{3}$ GWD, one can express any spinning CPW as a product of two AdS$_{2}$ GWD. A related question is to ask how the spinning OPE blocks in this case are related in terms of AdS$_{2}$ Geodesic operators. The main point of this note is to explore this connection between spinning OPE blocks and AdS$_{2}$ geodesic operators. \\

Our starting point is the observation that in 2D CFT, the method of finding an integral expression for scalar OPE block discussed in \cite{Ferrara:1971vh},\cite{Ferrara:1973vz}, can be easily extended to the case of spinning OPE block, constructed out of spinning primary field. This type of simple extension is special in CFT$_{2}$, where all spinning primaries are labelled by two real numbers($h$,$\bar{h}$), which is unavailable in higher dimensions.\footnote{In D=2, the conformal symmetry is infinite dimensional and so the OPE's can be organized into sum of blocks, governed by the full Virasoro symmetry, but here we consider only global OPE blocks, which are governed by the global conformal symmetry. In general, closed form expression of Virasoro conformal block is not known yet, except in a particular limit where the Virasoro block reduces to the global block \cite{Zamolodchikov:1985ie} and hence, one can use GWD prescription to compute the block from AdS$_{3}$ \cite{Fitzpatrick:2014vua}-\cite{Hijano:2015qja}.} For instance, using such simplification, available only in 2D, Osborn has derived \cite{Osborn:2012vt} the closed form expression of spinning conformal block (\ref{2dclosed}) by a simple generalization from the spinless case in CFT$_{2}$, without having to use the embedding space formalism , which is necessary for deriving expressions for conformal blocks for four point function of non-zero spin operators in higher dimensions.\\


In this note, we try to give a dual interpretation of the generalized integral formula for spinning OPE block in CFT$_{2}$. Interestingly, the integral expression contains the HKLL kernel \cite{Hamilton:2005ju},\cite{Hamilton:2006az} of AdS$_{2}$ fields. In particular, for the case when the OPE blocks are constructed out of conserved current primaries, one can use the HKLL representation  of scalar bulk fields in AdS$_{2}$ to recast this formula as a product of two geodesic integrals of AdS$_{2}$ bulk fields, along the lines of \cite{daCunha:2016crm}. To make contact with the bulk dual, we use the HKLL representation for massless  higher spin fields \cite{Sarkar:2014dma} in AdS$_3$. In \cite{deBoer:2016pqk}, \cite{Czech:2016tqr} and \cite{Dyer:2017zef}, bulk dual of spinning OPE block in CFT$_{d}$, has been discussed in terms of gauge invariant bulk quantity integrated over minimal surface. To the best of our knowledge, the bulk interpretation for conserved spinning OPE blocks in 2D, that we present in this note using special features of CFT$_{2}$, is different from those and is valid for any higher spin symmetric conserved currents. \\

The note is organised as follows. In the next section, we review the relevant parts of (\cite{Czech:2016xec} $\&$ \cite{daCunha:2016crm}) about the bulk duals of scalar OPE blocks as discussed therein.  In section (\ref{sec3}), we discuss specifically the case of the conserved current-spinning OPE blocks (OPE blocks built out of primaries which are conserved currents). We show that, their duals are given by AdS$_2$ geodesic operators and we also discuss their connection to AdS$_3$ massless higher spin fields using the results of \cite{Sarkar:2014dma}. The details of the calculations are given in two appendices. In section (\ref{sec4}), using this bulk AdS$_2$  representation for the conserved current OPE blocks and some known results of GWD, we compute the spinning conformal blocks and match with known results of Osborn \cite{Osborn:2012vt}.  We end up with a discussion of our results and some comments on possible future extension.

\section{Review of scalar geodesic operators}\label{sec2}
 
OPE blocks ($B_{k}^{ij}$) are defined as the contribution of a conformal family associated to a given primary field ($\mathcal{O}_k$) to the OPE of two primary operators($\mathcal{O}_i$, $\mathcal{O}_j$) of dimensions $\Delta_{i}$,$\Delta_{j}$ respectively. Mathematically, 

\begin{align}\label{opeblock}
\mathcal{O}_{i}(x_{1})\mathcal{O}_{j}(x_{2}) = |x_{12}|^{-(\Delta_{i}+\Delta_{j})}\sum_{k}C_{ijk}B_{k}^{ij}(x_{1},x_{2})
\end{align}
OPE blocks corresponding to non-spinning primary operators\footnote{ie when $h_k = \bar{h}_k$},  were identified with  the geodesic weighted integral of bulk scalar fields in AdS, independently in \cite{Czech:2016xec},\cite{deBoer:2016pqk} and \cite{daCunha:2016crm}. In the former two papers, this identification was achieved  by showing that, both are solutions to identical differential equations satisfying same boundary conditions, while in \cite{daCunha:2016crm} this was realized by recasting an old formula due to \cite{Ferrara:1971vh},\cite{Ferrara:1973vz} in terms of bulk variables.


The conformal Casimir equation satisfied by the OPE blocks, can be visualized geometrically by identifying them as local fields on an auxiliary space, the so-called `kinematic space'(k-space) made  of pair of points of the CFT \cite{Czech:2016xec},\cite{deBoer:2016pqk}. In particular, for 2D CFT,
the Casimir eigenvalue equation of the OPE block precisely gives the Klein-Gordon equation in kinematic space \footnote{ For the more general case with unequal weights($\Delta_i \neq \Delta_j$) as well as with unequal spin($l_i \neq l_j$), the modification of the Casimir equation will lead to a modified field equation in kinematic space. For more details, see the appendix \ref{transformation}.}.
\begin{align}\label{addition}
\Box_{K}B_{k} = C_{k}B_{k}; \quad C_{k}=-\Delta_{k}(\Delta_{k}-2)-l_{k}^{2}
\end{align} 

By its very definition, as the space of pair of points in CFT$_{2}$, the kinematic space can also be identified as the space of space-like geodesics in AdS$_{3}$ \cite{Czech:2015qta}, ending on the boundary. 
It turns out that, the dual of the OPE blocks are 'geodesic operators', which are weighted integrals of bulk fields over geodesics $\gamma$, with endpoints in the boundary $x_{1}$,$x_{2}$. 
\begin{align}
B_{k}(x_{1},x_{2}) \sim \int_{\gamma}ds\phi(x(s))
\end{align}

In \cite{Czech:2016xec}, using the techniques of integral transform(X-ray transform in particular), the authors have explicitly shown that, for a massive scalar field in AdS, the geodesic operator on the RHS, satisfies the differential equations 
\begin{align}
\int_{\gamma}(\Box_{AdS}-m^{2})\phi(x) = 0 \implies (\Box_{k}+m^{2})\int_{\gamma}\phi(x) = 0
\end{align}
where $m^{2} = -C_{k}$ with $l_{k} = 0$.\\
Finally, using the `AdS/CFT boundary condition' for bulk scalar field $\phi(y,z)$ in terms of the boundary scalar primary $\mathcal{O}(y)$, they showed that, the geodesic operator satisfies the same boundary condition as that of the OPE block at the coincidence limit of two operators \cite{Czech:2016xec}. So that, 
\begin{align}
\lim_{z\rightarrow 0}\phi_{AdS}(y,z=x_{1}-x_{2})\sim  |x_{1}-x_{2}|^{\Delta}\mathcal{O}_{\Delta}(y) =\lim_{x_{1}\rightarrow x_{2}}B_{\Delta}(x_{1},x_{2})
\end{align}

In \cite{daCunha:2016crm}, this description was arrived at independently from an old integral formula of OPE block of dimension $\Delta$, derived in \cite{Ferrara:1971vh},\cite{Ferrara:1973vz}, for the OPE of two spacelike separated scalar operators with dimensions $\Delta_{i}$,$\Delta_{j}$ in $CFT_d$. 
\begin{align}\label{guica}
A(x)B(0)|_{\Delta} \sim \frac{\mathcal{B}^{-1}_{ij}}{|x|^{\Delta_{i}+\Delta_{j}}}\int_{0}^{1}\frac{du}{u(1-u)}\left(\frac{u}{1-u}\right)^{\frac{\Delta_{ij}}{2}}\Gamma\left(\nu+1\right)2^{\nu} \times \nonumber \\
\int \frac{d^{d}p}{(2\pi)^{d}}\frac{e^{iup.x}}{\left(-p^{2}\right)^{\frac{\nu}{2}}}\left(u(1-u)x^{2}\right)^{\frac{d}{4}}\textit{J}_{\nu}\left(\sqrt{-u(1-u)x^{2}p^{2}}\right)\mathcal{O}(p)
\end{align}
where $\nu = \Delta-\frac{d}{2}$,$\Delta_{ij} = \Delta_{i}-\Delta_{j}$ and the Euler Beta function $\mathcal{B}_{ij} = \mathcal{B}_{ij}(\frac{\Delta+\Delta_{ij}}{2},\frac{\Delta-\Delta_{ij}}{2})$. The idea of \cite{daCunha:2016crm}, is to rewrite the integral using some change of variables, as a integral of local bulk field smeared along a geodesic, whose end-points are the two points of the OPE. The basic ingredient in this analysis is the result of HKLL,  which gives an integral expression of local free bulk scalar field in terms of boundary CFT operator.
\begin{align}\label{5}
\phi_{HKLL}^{(0)}(z,x)=2^{\nu}\Gamma(\nu +1)\int\frac{d^{d}p}{(2\pi)^{d}}\frac{e^{ip.x}}{(-p^{2})^{\frac{\nu}{2}}}z^{\frac{d}{2}}\textit{J}_{\nu}(z\sqrt{-p^{2}})\mathcal{O}(p)
\end{align}
Using the geodesic length parameter $\lambda$ and the result of HKLL, the OPE integral becomes,
\begin{align}
A(x)B(0)|_{\Delta} \sim \frac{\mathcal{B}^{-1}_{ij}}{|x|^{\Delta_{i}+\Delta_{j}}}\int_{-\infty}^{\infty}d\lambda e^{-\lambda\Delta_{AB}}\phi_{HKLL}^{(0)}(y(\lambda))
\end{align}

Also in kinematic space language, conformal blocks are just propagators in that space. In a similar fashion, the dual of CPW, that is GWD in the bulk, can be rederived, just from the more fundamental holographic identification of `OPE block' with `Geodesic operator'.

\section{Spinning OPE blocks in AdS$_{3}$/CFT$_{2}$}\label{sec3}
This section contains the main analysis and result of this note. Under conformal transformations, any primary tensor field $\phi_{z,\dots,z,\bar{z},\dots,\bar{z}}$ in CFT$_{2}$ of rank $\Delta$ transforms as:
\begin{align}
\phi_{z,\dots,z,\bar{z},\dots,\bar{z}}(z,\bar{z}) = \left(\frac{\partial f(z)}{\partial z}\right)^{h}\left(\frac{\partial \bar{f}(\bar{z})}{\partial \bar{z}}\right)^{\bar{h}}\phi_{z,\dots,z,\bar{z},\dots,\bar{z}}(f(z),\bar{f}(\bar{z}))
\end{align} 
Thus any quasi primary field with spin $l$ and conformal dimensions $(h,\bar{h})$, transforms like tensor of rank $h+\bar{h}=\Delta$, with $h$ of  $z$($\equiv x+t$) indices and $\bar{h}$ of $\bar{z}$($\equiv x-t$) indices. In 2D CFT, three point function of any  primary fields with arbitrary spin, has a simple structure, unlike in higher dimensions. Using the method of \cite{Ferrara:1971vh},\cite{Ferrara:1973vz}, we can derive an integral expression for the OPE of two arbitrary spin primaries. This leads to (see appendix \ref{integral expression}.) the following expression of OPE block of dimension $(h_{k},\bar{h}_{k})$ for two spinning operators $A(z_{1},\bar{z_{1}})$ and $B(0,0)$ of conformal dimension $h_{i}$ and $h_{j}$ respectively,
\begin{align}\label{4}
A(z_{1},\bar{z_{1}})B(0,0)|_{h_{k},\bar{h}_{k}} 
= \mathcal{B}^{-1}_{AB}\left(\frac{1}{z_{1}^{2}}\right)^{\frac{1}{2}(h_{i}+h_{j})}\int_{0}^{1}\frac{du}{u(1-u)}\left(\frac{u}{1-u}\right)^{\frac{h_{ij}}{2}}\Gamma\left(h_{k}+\frac{1}{2}\right)2^{h_{k}-\frac{1}{2}} \times\nonumber \\
\int \frac{dp}{2\pi}\frac{e^{iuz_{1}p}}{\left(-p^{2}\right)^{\frac{h_{k}}{2}-\frac{1}{4}}}\left(u(1-u)z_{1}^{2}\right)^{\frac{1}{4}}\textit{J}_{h_{k}-\frac{1}{2}}\left(\sqrt{-u(1-u)z_{1}^{2}p^{2}}\right) \times\nonumber \\
\mathcal{B}^{-1}_{\bar{A}\bar{B}}\left(\frac{1}{\bar{z}_{1}^{2}}\right)^{\frac{1}{2}(\bar{h}_{i}+\bar{h}_{j})}\int_{0}^{1}\frac{dv}{v(1-v)}\left(\frac{v}{1-v}\right)^{\frac{\bar{h}_{ij}}{2}}\Gamma\left(\bar{h}_{k}+\frac{1}{2}\right)2^{\bar{h}_{k}-\frac{1}{2}} \times\nonumber \\
\int \frac{dq}{2\pi}\frac{e^{iv\bar{z}_{1}q}}{\left(-q^{2}\right)^{\frac{\bar{h}_{C}}{2}-\frac{1}{4}}}\left(v(1-v)\bar{z}_{1}^{2}\right)^{\frac{1}{4}} \textit{J}_{\bar{h}_{k}-\frac{1}{2}}\left(\sqrt{-v(1-v)\bar{z}_{1}^{2}q^{2}}\right)C(p,q)
\end{align}

Our goal is to interpret this integral OPE formula in AdS$_{3}$/CFT$_{2}$ context on the lines of \cite{daCunha:2016crm}. In this section, we will restrict ourselves to the case, where the OPE block corresponds to a symmetric traceless conserved current in $CFT_2$. The main reason for this is that, for a general spin-s operator, there is no analogue of the HKKL construction in the literature.         

\subsection{symmetric traceless conserved currents in CFT$_{2}$}\label{symtraceless}

 A symmetric conserved current $J_{\mu_1,...\mu_l}$ in 2D CFT has only two non vanishing components, which are purely holomorphic or anti-holomorphic: $J(z)\equiv J_{z....z}(z)$ and $\bar{J}(\bar{z}) \equiv J_{\bar{z}...\bar{z}}(\bar{z})$. 
 Such a conserved primary current has to satisfy the condition 
$\Delta_{J}=|s|$ in two dimensions. This implies that the component $J$ has dimensions $(s,0)$ and the component $\bar{J}$ has dimensions$(0,s)$. 
We could consider OPE blocks, constructed out of such conserved primaries. A generic operator would be of the form $J_1(z)\bar{J}_2(\bar{z})$, constructed out of the holomorphic component of one current with dimension $h_{k}$ and the antiholomorphic component of another current with dimension $\bar{h}_{k}$. For these operators, $C(p,q)$ factorizes into a product. This simplifies (\ref{4})  and then using (\ref{5}) we get,
\begin{align}\label{ads2/cft1}
A(z_{1},\bar{z_{1}})B(0,0) \sim \int_{-\infty}^{\infty}d\lambda e^{-\lambda h_{AB}}\phi_{AdS_{2}}^{(0)}(x(\lambda))\int_{-\infty}^{\infty}d\lambda' e^{-\lambda'\bar{h}_{AB}}\phi_{AdS_{2}}^{(0)}(x'(\lambda'))
\end{align}

 Where, following \cite{daCunha:2016crm}, we have defined the bulk coordinate as, $y(u)=\sqrt{u(1-u)z_{1}^{2}}$, $y'(u)=\sqrt{v(1-v)\bar{z}_{1}^{2}}$ and the corresponding boundary coordinates $z_{1}(u)=uz_{1}$ and $\bar{z}_{1}(v)=v\bar{z}_{1}$ in Poincare AdS$_{2}$ and introduced the geodesic length parameter $\lambda$,$\lambda'$, such that, $u=\frac{1}{1+e^{2\lambda}}$ and $v=\frac{1}{1+e^{2\lambda'}}$.
Also, $x(\lambda) = (y(\lambda),z_{1}(\lambda))$ and $x'(\lambda') = (y'(\lambda'),\bar{z}_{1}(\lambda'))$ are the bulk points over geodesics, where the two AdS$_{2}$ fields live. Therefore, 
\begin{align}
\phi_{AdS_{2}}^{(0)}(z,x)\sim \int\frac{dp}{2\pi}K_{AdS_{2}}(z,x,p)\mathcal{O}(p);\quad K_{AdS_{2}}(z,x,p) = \frac{e^{ip.x}}{(-p^{2})^{\frac{h_{k}}{2}-\frac{1}{4}}}z^{\frac{1}{2}}\textit{J}_{h_{k}-\frac{1}{2}}(z\sqrt{-p^{2}}).
\end{align}
The masses of the corresponding AdS$_{2}$ fields are $h_{k}(h_{k}-1)$ and $\bar{h}_{k}(\bar{h}_{k}-1)$, where $h_k$ and $\bar{h}_k$ are the spins of the two conserved currents, out of which we have constructed the OPE block. 
The fields $\phi_{AdS_{2}}$ satisfies the AdS/CFT boundary condition,$\lim_{y\rightarrow 0}\phi_{AdS_{2}}(y,z_{1})=y^{h_{k}}\mathcal{O}(z_{1})$(similar boundary condition applies for the other dual operators of dimension $\bar{h}_{k}$ corresponding to the other AdS$_{2}$ field).  

\subsubsection{kinematic space description}\label{kinspacedes}

Following \cite{Czech:2016xec}, it is easy to find the equation, that the OPE blocks satisfy for the case, when $h_i\neq h_j\neq \bar{h}_i \neq \bar{h_j}$. 
\begin{equation}
[2(z_{1}-z_{2})^{2}\frac{\partial^{2}}{\partial z_{1} \partial z_{2}} + 2(h_{2}-h_{2})(z_{1}-z_{2})(\partial_{z_{1}}+\partial_{z_{2}}) +cc]B^{ij}_{k}= [2h_{k}(1-h_{k})+2\bar{h}_{k}(1-\bar{h}_{k})]B^{ij}_{k}
\end{equation}
The associated boundary conditions are \cite{Czech:2016xec}, 
\begin{equation}
\lim_{z_{1},\bar{z}_{1}\rightarrow z_{2},\bar{z}_{2}}B_{h_{k}}^{ij}(z_{1},z_{2};\bar{z}_{1},\bar{z}_{2})=(z_{1}-z_{2})^{h_{k}}(\bar{z}_{1}-\bar{z}_{2})^{\bar{h}_{k}}\mathcal{O}_{h_{k}}(z_{1})\mathcal{O}_{\bar{h}_{k}}(\bar{z}_{1}) 
\end{equation}
We can also show explicitly \footnote{The details of the calculations are given in appendix \ref{intert}.} that the RHS of (\ref{ads2/cft1}) satisfies the above equation as well as the above boundary condition.
\begin{align}
&[2(z_{1}-z_{2})^{2}\frac{\partial^{2}}{\partial z_{1} \partial z_{2}} + 2(\bar{z}_{1}-\bar{z}_{2})^{2}\frac{\partial^{2}}{\partial \bar{z}_{1} \partial \bar{z}_{2}} + 
2 h_{ij}(z_{1}-z_{2})(\partial_{z_{1}}+\partial_{z_{2}})+ \nonumber \\
&2 \bar{h}_{ij}(\bar{z}_{1}-\bar{z}_{2})(\partial_{\bar{z}_{1}}+\partial_{\bar{z}_{2}})] \int_{\gamma}ds e^{-sh_{ij}}\phi_{AdS_{2}}(x(s)) \int_{\gamma'}ds' e^{-s'\bar{h}_{ij}}\phi_{AdS_{2}}(x'(s'))  \nonumber \\
&= -\int_{\gamma}ds e^{-sh_{ij}}\int_{\gamma'}ds' e^{-s'\bar{h}_{ij}}[\phi'(x(s'))\Box_{AdS_{2}}\phi(x(s))+\phi(x(s))\Box_{\bar{AdS_{2}}}\phi'(x(s'))]
\end{align}
Also one can check that the initial condition of field in k-space agrees with the AdS/CFT boundary condition via such geodesic operator, in this case.
\begin{align}
&\lim_{z_{1},\bar{z}_{1}\rightarrow z_{2},\bar{z}_{2}}B_{h_{k}}^{ij}(z_{1},z_{2};\bar{z}_{1},\bar{z}_{2})= \lim_{z_{1},\bar{z}_{1}\rightarrow z_{2},\bar{z}_{2}}R\phi_{AdS_{2}}(z(s),z_{1}(s))R\phi_{AdS_{2}}(z'(s'),\bar{z}_{1}(s'))
\end{align}
where $R \equiv \int_{\gamma} e^{-sh_{ij}}ds$.
\subsubsection{higher spin fields in AdS$_3$}\label{massless}
Equation (\ref{ads2/cft1}) expresses the higher spin OPE block for symmetric traceless conserved currents in CFT$_{2}$ in terms of integrated scalar fields in $AdS_2$. Naively, this happens, because the components of these currents are purely holomorphic or anti-holomorphic and therefore they are thought of as scalar operators lived in a CFT$_1$, where the CFT$_1$ coordinate was $z, \bar{z}$ respectively. Thus it might be suggestive that those (anti)holomorphic blocks could be identified with scalar geodesic operators in the dual AdS$_2$ description. 

Nevertheless, we still need to understand what these fields correspond to, from the dual AdS$_3$ perspective. In AdS$_{3}$/CFT$_{2}$, the dual of such symmetric traceless conserved currents are massless higher spin fields. Such fields have two non vanishing components in AdS$_3$. We would like to understand how these are related to the two scalar fields in AdS$_2$. To do this, we will use the results of \cite{Sarkar:2014dma}, who have generalized the construction of \cite{Kabat:2012hp},\cite{Heemskerk:2012np} to the case of massless higher spin fields in AdS$_{d+1}$.

In AdS$_{d+1}$, the equation of motion for massless, totally symmetric, integer, higher spin rank-$l$ free field $\phi_{\mu_{1}\mu_{2}\dots\mu_{l}}$, is obtained by generalizing Fronsdal equation in AdS \cite{Vasiliev:1990en}-\cite{Mikhailov:2002bp}. 

\begin{align}
\nabla^{2}\phi_{\mu_{1}\mu_{2}\dots\mu_{l}}-l\nabla_{(\mu_{1}}\nabla^{\mu}\phi_{\mu_{2}\dots\mu_{l})\mu} +\frac{l(l-1)}{2}\nabla_{(\mu_{1}}\nabla_{\mu_{2}}\phi_{\mu_{3}\dots\mu_{l})\mu}^{\quad \quad \quad\mu} \nonumber \\
-((l-2)(l+d-2)-l)\phi_{\mu_{1}\mu_{2}\dots\mu_{l}} - \frac{l(l-1)}{4}g_{(\mu_{1}\mu_{2}}\phi_{\mu_{3}\dots\mu_{l})\mu}^{\quad \quad\quad\mu} =0
\end{align}
Imposing the holographic gauge fixing condition \footnote{in which all the holographic $y$ component of gauge fields is made to vanish.
\begin{align}
\phi_{y\dots y} = \phi_{\mu_{1}y\dots y} = \dots = \phi_{\mu_{1}\dots\mu_{l-1}y} = 0
\end{align}
}and by redefining $\psi_{\mu_{1}\dots\mu_{l}} = y^{l}\phi_{\mu_{1}\dots\mu_{l}}$, It was shown in \cite{Sarkar:2014dma}, that the higher spin equations reduce to a set of scalar equations:
\begin{align}\label{scalar equation}
\partial_{\alpha}\partial^{\alpha}\psi_{\mu_{1}\dots\mu_{l}}+y^{d-1}\partial_{y}(y^{1-d}\partial_{y}\psi_{\mu_{1}\dots\mu_{l}})-\frac{(l-2)(l+d-2)}{y^{2}}\psi_{\mu_{1}\dots\mu_{l}} = 0
\end{align}
Thus every components of $\psi_{\mu_{1}\dots\mu_{l}}$ satisfies scalar free equations with mass $m^{2} = (l-2)(l+d-2)$ and the boundary condition $\lim_{y\rightarrow 0}\psi_{\mu_{1}\dots\mu_{l}}(y,x)=y^{\Delta}\psi_{\mu_{1}\dots\mu_{l}}(x)$, where $\Delta = l+d-2$ corresponds to dimension of the conserved primary in the boundary. In this way one can construct higher spin bulk field in terms of scalar smearing function by smearing the boundary conserved higher spin operators over a complexified boundary. The smeared integral takes the form:
\begin{align}\label{smearing}
\psi_{\mu_{1}\dots\mu_{l}} = \frac{\Gamma(l+\frac{d}{2}-1)}{\pi^{\frac{d}{2}}\Gamma(l-1)} \int_{t'^{2}+|x'|^{2}<y^{2}}dt' d^{d-1}x' \left(\frac{y^{2}-t'^{2}-x'^{2}}{y}\right)^{l-2}\mathcal{O}_{\mu_{1}\dots\mu_{l}}(t+t',x+ix')
\end{align}
In our case, we are interested in spinning fields of AdS$_{3}$ with coordinates $y,z,\bar{z}$. 
The traceless and transverse condition of bulk gauge field implies, that only two independent components are non vanishing i.e $\psi_{zzz\dots}(z,y)$ and $\psi_{\bar{z}\bar{z}\bar{z}\dots}(\bar{z},y)$. Therefore from (\ref{smearing}) we have,
\begin{align}
\psi_{zzz\dots}(z,y)=\frac{l-1}{\pi} \int_{t'^{2}+y'^{2}<y^{2}}dt' dy' \left(\frac{y^{2}-t'^{2}-y'^{2}}{y}\right)^{l-2}\mathcal{O}_{zzz\dots}(t+t',x+iy')
\end{align}
where $(z,\bar{z}) = (x+t,x-t)$. As $\mathcal{O}_{zzz\dots}$ is only a function of $z$, using $t'=r\cos\theta$ and $y'=r\sin\theta$, the integral becomes,
\begin{align}
\psi_{zzz\dots}(z,y)=\frac{l-1}{\pi}\int_{0}^{y}rdr\left(\frac{y^{2}-r^{2}}{y}\right)^{l-2}\int_{0}^{2\pi}d\theta\mathcal{O}_{zzz\dots}(z+re^{i\theta})
\end{align}
Using $x=e^{i\theta}$ the $\theta$ integral reduces a contour integral inside a circle.
\begin{align}
\psi_{zzz\dots}(z,y)=\frac{l-1}{\pi}\int_{0}^{y}rdr\left(\frac{y^{2}-r^{2}}{y}\right)^{l-2}\oint \frac{\mathcal{O}_{zzz\dots}(z+rx)}{ix} dx
\end{align}
Analyticity of higher spin conserved currents indicates that the only pole at $x=0$ contributes in the $x$ integral and it gives $2\pi\mathcal{O}_{zzz\dots}(z)$. This further simplifies the $r$ integral as follows
\begin{align}
\psi_{zzz\dots}(z,y)=2(l-1)\mathcal{O}_{zzz\dots}(z)\int_{0}^{y}rdr\left(\frac{y^{2}-r^{2}}{y}\right)^{l-2}
\end{align}
It is now straightforward to see the final expression we get from this:
\begin{align}\label{psi}
\psi_{zzz\dots}(z,y)=y^{l}\mathcal{O}_{zzz\dots}(z) \quad \textrm{or,} \quad \phi_{zzz\dots}(z)=\mathcal{O}_{zzz\dots}(z)
\end{align}
In a similar fashion one can also show that the other non-vanishing component of higher spin fields i.e $\psi_{\bar{z}\bar{z}\bar{z}\dots}(\bar{z},y)$ simplifies.\footnote{Similar construction for linearized gravity in AdS$_{3}$ appears in \cite{Kabat:2012hp}.}
\begin{align}
\psi_{\bar{z}\bar{z}\bar{z}\dots}(\bar{z},y) = y^{l}\mathcal{O}_{\bar{z}\bar{z}\bar{z}\dots}(\bar{z})  \quad \textrm{or,} \quad \phi_{\bar{z}\bar{z}\bar{z}\dots}(\bar{z})=\mathcal{O}_{\bar{z}\bar{z}\bar{z}\dots}(z)
\end{align}
Therefore in the radial gauge, these two non-vanishing independent components of AdS$_{3}$ gauge fields are basically the (anti)holomorphic components of conserved currents live at the boundary. This simplification suggests a way to re-express AdS$_{2}$ scalar field in terms of the components of higher spin field in AdS$_{3}$. Following HKLL construction for scalar in AdS$_{2}$ \cite{Hamilton:2005ju},\cite{Hamilton:2006az} and using (\ref{psi}) we have,
\begin{align}
\phi_{AdS_{2}}^{(0)}(y,z) \sim \int_{-\infty}^{\infty} dz' \left(\frac{y^{2}-(z-z')^{2})}{y}\right)^{\Delta - 1} \theta(y-|z-z'|)\phi_{zzz\dots}(z')
\end{align}
Therefore, the connection between CFT$_{2}$ spinning OPE block and it's dual representation in terms of two copies of AdS$_{2}$ free scalar field, can be related to the AdS$_{3}$/CFT$_{2}$ correspondence for massless higher spin/conserved boundary current, following the construction for massless higher spin fields in \cite{Sarkar:2014dma} in a specific gauge. In a different gauge, the form of the explicit map would be different.

\section{Spinning conformal block and geodesic witten diagram}\label{sec4}
In this section we want to investigate the four-point spinning conformal partial wave in CFT$_{2}$, using the relation (\ref{ads2/cft1}) of boundary OPE in terms of bulk field. The conformal partial wave is defined by projection of conformal four point function onto a conformal family $\mathcal{O}_{k}$.
\begin{align}
W_{h_{k},\bar{h}_{k}}(z_{i},\bar{z}_{i}) = <A(z_{1},\bar{z}_{1})B(z_{2},\bar{z}_{2})\mathbb{P}_{\mathcal{O}}C(z_{3},\bar{z}_{3})D(z_{4},\bar{z}_{4})>
\end{align}
where $\mathbb{P}_{\mathcal{O}}$ is the projector onto the conformal family $\mathcal{O}$ of dimension $h_{k},\bar{h}_{k}$. Using (\ref{ads2/cft1}) we get, 
 
\begin{align}
W_{h_{k},\bar{h}_{k}} = \frac{\mathcal{B}^{-1}_{AB}\mathcal{B}^{-1}_{CD}}{z_{12}^{(h_{A}+h_{B})}z_{34}^{(h_{C}+h_{D})}}\int_{\gamma_{AB}}d\lambda e^{-\lambda h_{AB}} \int_{\gamma_{CD}}d\lambda'' e^{-\lambda'' h_{CD}} <\phi^{(0)}_{h_{k}}(y(\lambda))\phi^{(0)}_{h_{k}}(y''(\lambda''))> \times \nonumber \\
 \frac{\mathcal{B}^{-1}_{\bar{AB}}\mathcal{B}^{-1}_{\bar{CD}}}{\bar{z}_{12}^{(\bar{h}_{A}+\bar{h}_{B})}\bar{z}_{34}^{(h_{C}+h_{D})}}\int_{\gamma_{A'B'}}d\lambda' e^{-\lambda' \bar{h}_{AB}} \int_{\gamma_{C'D'}}d\lambda''' e^{-\lambda''' \bar{h}_{CD}}<\phi^{(0)}_{\bar{h}_{k}}(y'(\lambda'))\phi^{(0)}_{\bar{h}_{k}}(y'''(\lambda'''))>
\end{align}
Where $\phi_{h_{k}}^{(0)}$ denotes the free field in AdS$_{2}$ with corresponding mass $h_{k}(h_{k}-1)$. The bulk to boundary propagator for massive free scalar field in AdS$_{2}$ is $G_{b\partial}(y,z_{i}) = \left(\frac{y}{y^{2}+|z-z_{i}|^{2}}\right)^{h_{i}}$. Here, on the geodesic the bulk and boundary coordinates are parametrized by,
\begin{align}
y(u(\lambda))=\sqrt{u(1-u)(z_{1}-z_{2})^{2}} \quad \textrm{and} \quad z_{i}(u(\lambda)) = z_{2} + u(z_{1}-z_{2})
\end{align}
where $u(\lambda) = \frac{1}{1+e^{2\lambda}}$. Using these variables, the bulk-boundary propagator reduces to,
\begin{align}
G_{b\partial}(y,z_{1}) = \frac{e^{-\lambda h_{A}}}{z_{12}^{h_{A}}} \quad , \quad G_{b\partial}(y,z_{2}) = \frac{e^{\lambda h_{B}}}{z_{12}^{h_{B}}}
\end{align}
Also the bulk two point function is the bulk to bulk propagator. Using these results, we get,
\begin{align}
&W_{h_{k},\bar{h}_{k}} =\mathcal{B}^{-1}_{AB}\mathcal{B}^{-1}_{CD} [\int_{\gamma_{AB}}d\lambda  \int_{\gamma_{CD}}d\lambda''  G_{b\partial}(y(\lambda),z_{1}) G_{b\partial}(y(\lambda),z_{2})   G_{b\partial}(y''(\lambda''),z_{3})  G_{b\partial}(y(\lambda''),z_{4})G_{bb}(y(\lambda),y''(\lambda''),h_{k})] \times \nonumber \\
 &\mathcal{B}^{-1}_{\bar{AB}}\mathcal{B}^{-1}_{\bar{CD}}[\int_{\gamma_{A'B'}}d\lambda' \int_{\gamma_{C'D'}}d\lambda'''G_{b\partial}(y'(\lambda'),\bar{z}_{1})G_{b\partial}(y'(\lambda'),\bar{z}_{2}) G_{b\partial}(y'''(\lambda'''),\bar{z}_{3})G_{b\partial}(y'''(\lambda'''),\bar{z}_{4})G_{bb}(y'(\lambda'),y'''(\lambda'''),\bar{h}_{k})]
\end{align}
Therefore, it reduces to the product of two GWD on AdS$_{2}$. This implies that, 2D spinning conformal block decouples into two CFT$_{1}$ conformal block. To be more specific, using GWD, it has been checked \cite{daCunha:2016crm} that the first part of the product precisely a CFT$_{1}$ conformal block $k_{h}(z)$,, $z=\frac{z_{12}z_{34}}{z_{13}z_{24}}$ \cite{Dolan:2011dv}. 
Thus, finally we get what we expect,
\begin{align}\label{conformal block}
W_{h_{k},\bar{h}_{k}}(z_{i},\bar{z}_{i})= z^{h_{k}} {}_2F_1\left(h_{k}-h_{12},h_{k}+h_{34};2h_{k};z\right)\bar{z}^{\bar{h}_{k}} {}_2F_1\left(\bar{h}_{k}-\bar{h}_{12},\bar{h}_{k}+\bar{h}_{34};2\bar{h}_{k};\bar{z}\right)
\end{align}

The above expression was arrived at by restricting to the OPE block, obtained from the primary $J_{h_k}(z)\bar{J}_{\bar{h}_k}(\bar{z})$. We can similarly work out the contribution from the OPE block, constructed out of $J_{\bar{h}_k}(z)\bar{J}_{h_k}(\bar{z})$. This will be given by: 

\begin{align}
W_{\bar{h}_{k},h_{k}}(z_{i},\bar{z}_{i})= z^{\bar{h}_{k}} {}_2F_1\left(\bar{h}_{k}-h_{12},\bar{h}_{k}+h_{34};2\bar{h}_{k};z\right)\bar{z}^{h_{k}} {}_2F_1\left(h_{k}-\bar{h}_{12},h_{k}+\bar{h}_{34};2h_{k};\bar{z}\right)
\end{align}
The full four point function would have contributions from both these conformal blocks as in (\ref{2dclosed}). Therefore, this result provides a consistency check for our identification of OPE block with AdS$_{2}$ geodesic operators (\ref{ads2/cft1}).

\section{Discussion}\label{sec6}

In this note, following the work of \cite{Ferrara:1971vh},\cite{Ferrara:1973vz}, we have derived an OPE block integral formula  for conserved currents and we have reinterpreted this formula along the lines of \cite{daCunha:2016crm}, as an expression in the bulk AdS. The corresponding OPE block constructed out of conserved current components $J_{i}(z)\bar{J}_{j}(\bar{z})$ can be expressed as a product of two geodesic integrals of scalars in $AdS_2$. It should be possible to reproduce the scalar field OPE block case (\ref{guica}), as a special case of our formula (\ref{4}) when $h=\bar{h}$, however it is proved to be technically hard at present.

We then showed that, how these scalar fields are related to the components of massless higher spin fields in $AdS_3$, following the work of \cite{Sarkar:2014dma}. So basically, this note emphasizes the interplay between HKLL and OPE block-geodesic operator story in a different way. As a consistency check of our expression (\ref{4}), we showed how this expression for conserved currents OPE block, leads to the correct expression for spinning conformal block using GWD prescription for CFT$_{1}$ conformal block. 

Apart from this integral representation (\ref{4}) in Fourier space, another integral expression (\ref{expression of B}) coming from Shadow operator formalism in position space for spinning OPE block, exists in the literature. The powerful feature of (\ref{4}) is the fact that, it contains the well-known HKLL kernel of AdS$_{2}$ scalar field. To see the connection of conserved current block with AdS$_{2}$ scalar field is an interesting problem from the expression of (\ref{expression of B}).

A special example of our formula is the symmetric combination of stress tensor OPE block($B_{T}$,$B_{\bar{T}}$), which corresponds to the modular Hamiltonian \cite{Czech:2016xec}. In our formalism, each stress tensor block can be represented by geodesic integral of AdS$_{2}$ scalar field with mass $m^{2} = 2$. On the other hand, in \cite{Czech:2016tqr}, the bulk dual of modular Hamiltonian is described as the fluctuation in the area of minimal surface. It would be nice to understand how to relate our construction with their result.

We already stated earlier that there is a general prescription to study spinning OPE blocks in any dimension in \cite{deBoer:2016pqk}, \cite{Czech:2016tqr} and \cite{Dyer:2017zef}. In particular, \cite{deBoer:2016pqk} describes how to deal with spin one blocks in terms of geodesic operator of dual $(d-1)$ form $\ast F$, while \cite{Czech:2016tqr} deals with spin two bulk field as we mentioned above. Also \cite{Dyer:2017zef} \footnote{see also \cite{Castro:2017hpx}.} gives a general procedure to study spinning geodesic Witten diagram(as well as geodesic operator) by integrating spinning vertex over geodesic. Thus, these works are inherently deal with gauge invariant bulk observables. On the other hand, our work connects to traditional HKLL program of reconstructing bulk gauge fields from the boundary currents by choosing a specific gauge. Apart from giving different representation of spinning OPE blocks, it provides an example how components of gauge field in AdS$_{3}$ can be related to AdS$_{2}$ scalars. The feature of (anti)holomorphic factorization of conformal blocks and conserved current OPE blocks in CFT$_{2}$, is not at all obvious from bulk point of view. Because there is no direct connection of bulk dynamics between AdS$_{3}$ and AdS$_{2}$ a priori. Furthermore, the presence of AdS$_{2}$ kernel in general spinning block formula (\ref{4}) indicates a much more general setting to explore this connection for further study. More precisely, it would be nice to do a repackaging of the formula into a single integration over AdS$_{3}$, to understand the picture more clearly for general non-conserved spinning blocks. However, since we do not have a HKLL construction for general massive higher spin fields, we are unable to relate this expression to fields in $AdS_3$. We do not therefore have much to say about this general case. It would be nice if one could make progress in this direction. Furthermore, it would be nice, if, one could extend this direction using embedding space formalism to study OPE blocks and their dual in higher dimensions.

\vspace*{1ex}
\noindent{\bf Acknowledgment:} 
I am grateful to Bobby Ezhuthachan for his constant support and guidance through out the entire course of this project and for clarifying some issues related to massless higher-spin fields in AdS. I would also like to thank him for carefully going through the manuscript and for suggesting several changes therein. My work is supported by a Senior Research Fellowship from CSIR.

\appendix
\section{Integral expression of spinning OPE blocks in CFT$_{2}$}\label{integral expression}
Here, we follow the similar path of \cite{Ferrara:1971vh},\cite{Ferrara:1973vz}, to obtain an integral expression for spinning OPE block, comes from OPE of two arbitrary spin primaries in CFT$_{2}$. In 2D CFT three point function of three spinning operators $C(z_{2},\bar{z_{2}}),A(z_{1},\bar{z_{1}})$ and $B(0,0)$ is given by
\begin{align}\label{1}
&<0|C(z_{2},\bar{z_{2}})A(z_{1},\bar{z_{1}})B(0,0)|0> 
= c_{ABC}\left(\frac{1}{(z_{2}-z_{1})^{2}}\right)^{\frac{1}{2}(h_{C}+h_{A}-h_{B})}\left(\frac{1}{z_{1}^{2}}\right)^{\frac{1}{2}(h_{A}+h_{B}-h_{C})} \times\nonumber \\
&\left(\frac{1}{z_{2}^{2}}\right)^{\frac{1}{2}(h_{B}+h_{C}-h_{A})}
\left(\frac{1}{(\bar{z}_{2}-\bar{z}_{1})^{2}}\right)^{\frac{1}{2}(\bar{h}_{C}+\bar{h}_{A}-\bar{h}_{B})}\left(\frac{1}{\bar{z}_{1}^{2}}\right)^{\frac{1}{2}(\bar{h}_{A}+\bar{h}_{B}-\bar{h}_{C})}\left(\frac{1}{\bar{z}_{2}^{2}}\right)^{\frac{1}{2}(\bar{h}_{B}+\bar{h}_{C}-\bar{h}_{A})}
\end{align}
Using the famous Feynman parametrization formula of $\frac{1}{A^{\alpha_{1}}B^{\alpha_{2}}} = \frac{\Gamma(\alpha_{1}+\alpha_{2})}{\Gamma(\alpha_{1})\Gamma(\alpha_{2})}\int_{0}^{1}du \frac{u^{\alpha_{1}-1}(1-u)^{\alpha_{2}-1}}{(uA+(1-u)B)^{\alpha_{1}+\alpha_{2}}}$, one can write the following \footnote{This formula holds for any complex $A$ and $B$, unless it contains zero in their convex hull. Here we choose $z$,$\bar{z}$ to be real in Lorentzian coordinate such that, $z=x+t,\bar{z}=x-t$. Thus only spacelike separated points are allowed in CFT.}
\begin{align}\label{2}
&\left(\frac{1}{(z_{2}-z_{1})^{2}}\right)^{\frac{1}{2}(h_{C}+h_{A}-h_{B})}\left(\frac{1}{z_{2}^{2}}\right)^{\frac{1}{2}(h_{B}+h_{C}-h_{A})}\left(\frac{1}{(\bar{z}_{2}-\bar{z}_{1})^{2}}\right)^{\frac{1}{2}(\bar{h}_{C}+\bar{h}_{A}-\bar{h}_{B})}\left(\frac{1}{\bar{z}_{2}^{2}}\right)^{\frac{1}{2}(\bar{h}_{B}+\bar{h}_{C}-\bar{h}_{A})} \nonumber \\
&\propto \int_{0}^{1}du u^{\frac{1}{2}(h_{C}+h_{A}-h_{B})-1} (1-u)^{\frac{1}{2}(h_{B}+h_{C}-h_{A})-1}\left[(z_{2} - uz_{1})^{2}\right]^{-h_{C}}\left[1+z_{1}^{2}\frac{u(1-u)}{(z_{2}-z_{1})^{2}}\right]^{-h_{C}} \times \nonumber \\
&\int_{0}^{1}dv v^{\frac{1}{2}(\bar{h}_{C}+\bar{h}_{A}-\bar{h}_{B})-1} (1-v)^{\frac{1}{2}(\bar{h}_{B}+\bar{h}_{C}-\bar{h}_{A})-1}\left[(\bar{z}_{2} - v\bar{z}_{1})^{2}\right]^{-\bar{h}_{C}}\left[1+\bar{z}_{1}^{2}\frac{v(1-v)}{(\bar{z}_{2}-\bar{z}_{1})^{2}}\right]^{-\bar{h}_{C}}
\end{align}
Expanding $\left[1+z_{1}^{2}\frac{u(1-u)}{(z_{2}-z_{1})^{2}}\right]^{-h_{C}}$ and $\left[1+\bar{z}_{1}^{2}\frac{v(1-v)}{(\bar{z}_{2}-\bar{z}_{1})^{2}}\right]$ in binomial series, (\ref{2}) becomes,
\begin{align}
 &=\int_{0}^{1}du u^{\frac{1}{2}(h_{C}+h_{A}-h_{B})-1} (1-u)^{\frac{1}{2}(h_{B}+h_{C}-h_{A})-1}\sum_{n=0}^{\infty}\frac{1}{n!}\frac{\Gamma(h_{C}+n)}{\Gamma(h_{C})}(-z_{1}^{2}u(1-u))^{n}\left[\frac{1}{(z_{2}-uz_{1})^{2}}\right]^{h_{C}+n} \times \nonumber \\
&\int_{0}^{1}dv v^{\frac{1}{2}(\bar{h}_{C}+\bar{h}_{A}-\bar{h}_{B})-1} (1-v)^{\frac{1}{2}(\bar{h}_{B}+\bar{h}_{C}-\bar{h}_{A})-1}\sum_{m=0}^{\infty}\frac{1}{m!}\frac{\Gamma(\bar{h}_{C}+m)}{\Gamma(\bar{h}_{C})}(-\bar{z}_{1}^{2}v(1-v))^{m}\left[\frac{1}{(\bar{z}_{2}-v\bar{z}_{1})^{2}}\right]^{\bar{h}_{C}+m} 
\end{align}
Let us consider the following quantity:
\begin{align}
&<0|\partial_{z^{2}_{1}}^{n}\partial_{\bar{z}^{2}_{1}}^{m}C(z_{2},\bar{z_{2}})C(z_{1},\bar{z_{1}}) |0> =  \partial_{z^{2}_{1}}^{n}\left[\frac{1}{(z_{2}-z_{1})^{2}}\right]^{h_{C}}\partial_{\bar{z}^{2}_{1}}^{m}\left[\frac{1}{(\bar{z}_{2}-\bar{z}_{1})^{2}}\right]^{\bar{h}_{C}} \nonumber \\
&= 4^{n}\frac{\Gamma(h_{C}+n)\Gamma(h_{C}+n+\frac{1}{2})}{\Gamma(h_{C})\Gamma(h_{C}+\frac{1}{2})}\left[\frac{1}{(z_{2}-z_{1})^{2}}\right]^{h_{C}+n}4^{m}\frac{\Gamma(\bar{h}_{C}+n)\Gamma(\bar{h}_{C}+m+\frac{1}{2})}{\Gamma(\bar{h}_{C})\Gamma(\bar{h}_{C}+\frac{1}{2})}\left[\frac{1}{(\bar{z}_{2}-\bar{z}_{1})^{2}}\right]^{\bar{h}_{C}+m}
\end{align}

Now using the identity $\left[\frac{1}{(z_{2}-uz_{1})^{2}}\right]^{h_{C}+n} = \exp[uz_{1}.\partial_{z}]\left[\frac{1}{(z_{2}-z)^{2}}\right]^{h_{C}+n}|_{z=0}$, we have
\begin{align}
&\left[\frac{1}{(z_{2}-uz_{1})^{2}}\right]^{h_{C}+n}\left[\frac{1}{(\bar{z}_{2}-v\bar{z}_{1})^{2}}\right]^{\bar{h}_{C}+m} =  \exp[uz_{1}.\partial_{z}]\exp[v\bar{z}_{1}.\partial_{\bar{z}}]\left(\frac{1}{4}\right)^{n}\left(\frac{1}{4}\right)^{m} \times \nonumber \\ 
&\frac{\Gamma(h_{C})\Gamma(h_{C}+\frac{1}{2})}{\Gamma(h_{C}+n)\Gamma(h_{C}+n+\frac{1}{2})}\frac{\Gamma(\bar{h}_{C})\Gamma(\bar{h}_{C}+\frac{1}{2})}{\Gamma(\bar{h}_{C}+m)\Gamma(\bar{h}_{C}+m+\frac{1}{2})} 
 <0|C(z_{2},\bar{z_{2}})\partial_{z^{2}}^{m}\partial_{\bar{z}^{2}}^{n}C(0,0) |0>
\end{align}
Putting this back into (\ref{2}), we get the following expression for three point function
\begin{align}\label{3}
&<0|C(z_{2},\bar{z_{2}})A(z_{1},\bar{z_{1}})B(0,0)|0> = c_{ABC}\mathcal{B}^{-1}_{AB}\mathcal{B}^{-1}_{\bar{A}\bar{B}}\left(\frac{1}{z_{1}^{2}}\right)^{\frac{1}{2}(h_{A}+h_{B}-h_{C})}\left(\frac{1}{\bar{z}_{1}^{2}}\right)^{\frac{1}{2}(h_{A}+h_{B}-h_{C})} \times \nonumber \\
& \int_{0}^{1}du u^{\frac{1}{2}(h_{C}+h_{AB})-1} (1-u)^{\frac{1}{2}(h_{C}-h_{AB})-1}\sum_{n=0}^{\infty}\frac{1}{n!}\left(\frac{1}{4}\right)^{n}\frac{\Gamma(h_{C}+\frac{1}{2})}{\Gamma(h_{C}+n+\frac{1}{2})}\left(-z_{1}^{2}u(1-u)\right)^{n} \times \nonumber \\
&\int_{0}^{1}dv v^{\frac{1}{2}(\bar{h}_{C}+\bar{h}_{A}-\bar{h}_{B})-1} (1-v)^{\frac{1}{2}(\bar{h}_{B}+\bar{h}_{C}-\bar{h}_{A})-1}\sum_{m=0}^{\infty}\frac{1}{m!}\left(\frac{1}{4}\right)^{m}\frac{\Gamma(\bar{h}_{C}+\frac{1}{2})}{\Gamma(\bar{h}_{C}+m+\frac{1}{2})}(-\bar{z}_{1}^{2}v(1-v))^{m} \times\nonumber \\
& \exp[uz_{1}.\partial_{z}]\exp[v\bar{z}_{1}.\partial_{\bar{z}}] <0|C(z_{2},\bar{z_{2}})\partial_{z^{2}}^{n}\partial_{\bar{z}^{2}}^{m}C(0,0) |0>
\end{align}
where $\mathcal{B}^{-1}_{\bar{A}\bar{B}} = \mathcal{B}^{-1}(\frac{h_{C}+h_{AB}}{2},\frac{h_{C}-h_{AB}}{2})$ is the Euler Beta function and $h_{AB} = h_{A}-h_{B}$. One can identify the term $\sum_{n=0}^{\infty}\frac{1}{n!}\left(\frac{1}{4}\right)^{n}\frac{\Gamma(h_{C}+\frac{1}{2})}{\Gamma(h_{C}+n+\frac{1}{2})}\left(-z_{1}^{2}u(1-u)\partial_{z^{2}}\right)^{n}$ to the Hypergeometric function ${}_0F_1\left(h_{C}+\frac{1}{2} ; -\frac{z_{1}^{2}}{4}u(1-u)\partial_{z^{2}}\right)$.
 Hence, from (\ref{3}), we can get the following integral expression for the contribution of the conformal family of a primary operator $C$ of dimension $h_{C},\bar{h}_{C}$ to the OPE of $A(z_{1},\bar{z}_{1})B(0,0)$ .

\begin{align}
&A(z_{1},\bar{z_{1}})B(0,0)|_{h_{C},\bar{h}_{C}}= c_{ABC}\mathcal{B}^{-1}_{AB}\mathcal{B}^{-1}_{\bar{A}\bar{B}}\left(\frac{1}{z_{1}^{2}}\right)^{\frac{1}{2}(h_{A}+h_{B}-h_{C})}\left(\frac{1}{\bar{z}_{1}^{2}}\right)^{\frac{1}{2}(\bar{h}_{A}+\bar{h}_{B}-\bar{h}_{C})} \times \nonumber \\
&\int_{0}^{1}du u^{\frac{1}{2}(h_{C}+h_{AB})-1} (1-u)^{\frac{1}{2}(h_{C}-h_{AB})-1} \int_{0}^{1}dv v^{\frac{1}{2}(\bar{h}_{C}+\bar{h}_{A}-\bar{h}_{B})-1} (1-v)^{\frac{1}{2}(\bar{h}_{B}+\bar{h}_{C}-\bar{h}_{A})-1} \times \nonumber \\
&{}_0F_1\left(h_{C}+\frac{1}{2} ; -\frac{z_{1}^{2}}{4}u(1-u)\partial_{z^{2}}\right) {}_0F_1\left(\bar{h}_{C}+\frac{1}{2} ; -\frac{\bar{z}_{1}^{2}}{4}v(1-v)\partial_{\bar{z}^{2}}\right)\exp[uz_{1}.\partial_{z}]\exp[v\bar{z}_{1}.\partial_{\bar{z}}]C(0,0) 
\end{align}
Using the definition of Bessel function in terms of hypergeometric function one can rewrite the following,
\begin{align}
&{}_0F_1\left(h_{C}+\frac{1}{2} ; -\frac{z_{1}^{2}}{4}u(1-u)\partial_{z^{2}}\right) = \left(\frac{\sqrt{u(1-u)z_{1}^{2}\partial_{z^{2}}}}{2}\right)^{\frac{1}{2}-h_{C}}\Gamma\left(h_{C}+\frac{1}{2}\right)\textit{J}_{h_{C}-\frac{1}{2}}\left(\sqrt{u(1-u)z_{1}^{2}\partial_{z^{2}}}\right)
\end{align}
After simplifying this expression and performing Fourier transform, we finally get the following form,

\begin{align}
&A(z_{1},\bar{z_{1}})B(0,0)|_{h_{C},\bar{h}_{C}} \nonumber \\
&= \mathcal{B}^{-1}_{AB}\left(\frac{1}{z_{1}^{2}}\right)^{\frac{1}{2}(h_{A}+h_{B})}\int_{0}^{1}\frac{du}{u(1-u)}\left(\frac{u}{1-u}\right)^{\frac{h_{AB}}{2}}\Gamma\left(h_{C}+\frac{1}{2}\right)2^{h_{C}-\frac{1}{2}} \nonumber \\
&\int \frac{dp}{2\pi}\frac{e^{iuz_{1}p}}{\left(-p^{2}\right)^{\frac{h_{C}}{2}-\frac{1}{4}}}\left(u(1-u)z_{1}^{2}\right)^{\frac{1}{4}}\textit{J}_{h_{C}-\frac{1}{2}}\left(\sqrt{-u(1-u)z_{1}^{2}p^{2}}\right) \nonumber \\
&\mathcal{B}^{-1}_{\bar{A}\bar{B}}\left(\frac{1}{\bar{z}_{1}^{2}}\right)^{\frac{1}{2}(\bar{h}_{A}+\bar{h}_{B})}\int_{0}^{1}\frac{dv}{v(1-v)}\left(\frac{v}{1-v}\right)^{\frac{\bar{h}_{AB}}{2}}\Gamma\left(\bar{h}_{C}+\frac{1}{2}\right)2^{\bar{h}_{C}-\frac{1}{2}} \nonumber \\
&\int \frac{dq}{2\pi}\frac{e^{iv\bar{z}_{1}q}}{\left(-q^{2}\right)^{\frac{\bar{h}_{C}}{2}-\frac{1}{4}}}\left(v(1-v)\bar{z}_{1}^{2}\right)^{\frac{1}{4}} \textit{J}_{\bar{h}_{C}-\frac{1}{2}}\left(\sqrt{-v(1-v)\bar{z}_{1}^{2}q^{2}}\right)C(p,q)
\end{align}

\section{Transformation properties of spinning OPE blocks and different integral representation}\label{transformation}
We will now see how an OPE block(from an OPE of two arbitrary spinning operators) in CFT$_{2}$ with arbitrary spin, changes under conformal transformation.
\begin{align}\label{one}
&B'_{k}(z_{1},\bar{z_{1}};z_{2},\bar{z_{2}})- B_{k}(z_{1},\bar{z_{1}};z_{2},\bar{z_{2}}) = \mathcal{L}_{B}B_{k}(z_{1},\bar{z_{1}};z_{2},\bar{z_{2}}) 
\end{align}
The OPE of two operators $A(z_{1}),B(z_{2})$ with conformal dimension $h_{1},h_{2}$, can be written in terms of OPE block $B_{k}$, as,
\begin{align}\label{two}
&A(z_{1},\bar{z_{1}})B(z_{2},\bar{z_{2}}) = (z_{1}-z_{2})^{-h_{1}-h_{2}}(\bar{z}_{1}-\bar{z}_{2})^{-\bar{h}_{1}-\bar{h}_{2}}B_{k}(z_{1},\bar{z_{1}};z_{2},\bar{z_{2}})
\end{align} 
Now for arbitrary local infinitesimal conformal transformation $z' \rightarrow z+\epsilon(z)$, we have,
\begin{align}\label{six}
&A'(z_{1},\bar{z_{1}})B'(z_{2},\bar{z_{2}})-A(z_{1},\bar{z_{1}})B(z_{2},\bar{z_{2}})\nonumber \\
&=[-\epsilon(z_{1})\partial_{z_{1}}-h_{1}\partial_{z_{1}}\epsilon(z_{1})- \bar{\epsilon}(\bar{z}_{1})\partial_{\bar{z}_{1}}-\bar{h}_{1}\partial_{\bar{z}_{1}}\bar{\epsilon}(\bar{z}_{1}) - (z_{1} \leftrightarrow z_{2})]A(z_{1},\bar{z_{1}})B(z_{2},\bar{z_{2}})
\end{align}
This leads to a transformation rule for OPE block as in (\ref{one}). We will now check this transformation property of OPE block for different symmetry generators in CFT$_{2}$. At last, we want to find a Casimir equation acting on OPE block. The general strategy is to rewrite the generators, acting on OPE block, as the spin zero generators acting on that. 

For translation $\mathcal{L}_{-1} = -i\partial_{z}$, we get,
\begin{align}
&\mathcal{L}_{-1}B_{k}= [-\hat{\mathcal{L}}_{-1}^{(0)}+(h_{1}+h_{2})(z_{1}-z_{2})^{(h_{1}+h_{2}-1)}(\bar{z}_{1}-\bar{z}_{2})^{\bar{h}_{1}+\bar{h}_{2}} \nonumber \\ 
&-(h_{1}+h_{2})(z_{1}-z_{2})^{(h_{1}+h_{2}-1)}(\bar{z}_{1}-\bar{z}_{2})^{\bar{h}_{1}+\bar{h}_{2}} +cc]B_{k} = -\hat{\mathcal{L}}_{-1}^{(0)}B_{k}
\end{align}
Where $-\hat{\mathcal{L}}_{-1}^{(0)}$ is the generator of translation for spin zero operators. Similarly for scaling we would have $\mathcal{L}_{0} = h+z\partial_{z}$, $\mathcal{L}_{0}B_{k} =-\hat{\mathcal{L}}_{0}^{(0)}B_{k}$ and for special conformal transformation, we would have $\mathcal{L}_{1} = i(2zh+z^{2}\partial_{z})$ and $\mathcal{L}_{1}B_{k}= [-\hat{\mathcal{L}}_{1}^{(0)} + (h_{1}-h_{2})(z_{1}-z_{2})+cc]B_{k}$. Where $-\hat{\mathcal{L}}_{0}^{(0)}$ and $-\hat{\mathcal{L}}_{1}^{(0)}$ are the generators of zero spin operators for scaling and special conformal transformation respectively.

Hence, the Conformal Casimir equation implies,
\begin{align}
&\mathcal{L}^{2}_{B}B_{k} = (-2\mathcal{L}^{2}_{0}+\mathcal{L}_{1}\mathcal{L}_{-1}+\mathcal{L}_{-1}\mathcal{L}_{1})B_{k} = C_{k}B_{k}\nonumber \\
&=(-2\hat{\mathcal{L}}^{(0)2}_{0}+\hat{\mathcal{L}}^{(0)}_{1}\hat{\mathcal{L}}^{(0)}_{-1} +\hat{\mathcal{L}}^{(0)}_{-1}\hat{\mathcal{L}}^{(0)}_{1})B_{k} + [2(h_{1}-h_{2})(z_{1}-z_{2})(\partial_{z_{1}}+\partial_{z_{2}}) +cc]B_{k} \nonumber \\
&=[2(z_{1}-z_{2})^{2}\frac{\partial^{2}}{\partial z_{1} \partial z_{2}} + 2(h_{1}-h_{2})(z_{1}-z_{2})(\partial_{z_{1}}+\partial_{z_{2}}) +cc]B_{k}
\end{align}
where $C_{k}=2h_{k}(1-h_{k})+2\bar{h}_{k}(1-\bar{h}_{k})$.
Therefore we showed that instead of $\hat{\mathcal{L}}^{(0)2}_{B}$, the Casimir gets correction terms $[2(h_{1}-h_{2})(z_{1}-z_{2})(\partial_{z_{1}}+\partial_{z_{2}}) +cc]$, due to the spin difference of two external operators, $h_{1} \neq h_{2}$.

In \cite{Czech:2016xec}, an expression for OPE block for same operator dimension and same spin is given, using Shadow operator formalism\cite{Ferrara:1973vz},\cite{Ferrara:1972xe}-\cite{Ferrara:1972uq},\cite{SimmonsDuffin:2012uy}. One can generalize that expression for unequal spin cases using similar path. The formal expression is,
\begin{align}
B_{k}(z_{1},\bar{z_{1}};z_{2},\bar{z_{2}}) = n_{ijk}(z_{1}-z_{2})^{h_{i}+h_{j}}(\bar{z}_{1}-\bar{z}_{2})^{\bar{h}_{i}+\bar{h}_{j}}\int_{z_{1}}^{z_{2}}d\omega \int_{\bar{z}_{1}}^{\bar{z}_{2}}d\bar{\omega}<\mathcal{O}_{i}(z_{1},\bar{z}_{2}) \mathcal{O}_{j}(z_{2},\bar{z}_{2}) \mathcal{\tilde{O}}_{k}(\omega,\bar{\omega})>\mathcal{O}_{k}(\omega,\bar{\omega})
\end{align}
where $\mathcal{\tilde{O}}_{k}(\omega,\bar{\omega})$ is the shadow operator with conformal dimension $\tilde{h}_{k}=\frac{d}{2}-h_{k}$. After plugging back the conformal three point function we finally get,
\begin{align}\label{expression of B}
B_{k}(z_{1},\bar{z_{1}};z_{2},\bar{z_{2}}) =& n_{ijk}\int_{z_{1}}^{z_{2}}d\omega \int_{\bar{z}_{1}}^{\bar{z}_{2}}d\bar{\omega}\left(\frac{(\omega-z_{1})(z_{2}-\omega)}{z_{2}-z_{1}}\right)^{h_{k}-1}\left(\frac{z_{2}-\omega}{\omega-z_{1}}\right)^{h_{ij}}\times \nonumber \\
&\left(\frac{(\bar{\omega}-\bar{z}_{1})(\bar{z}_{2}-\bar{\omega})}{\bar{z}_{2}-\bar{z}_{1}}\right)^{\bar{h}_{k}-1}\left(\frac{\bar{z}_{2}-\bar{\omega}}{\bar{\omega}-\bar{z}_{1}}\right)^{\bar{h}_{ij}}\mathcal{O}_{k}(\omega,\bar{\omega})
\end{align}
It is easy to verify that (\ref{expression of B}) satisfies the conformal Casimir equation by acting the Casimir operator $[2(z_{1}-z_{2})^{2}\frac{\partial^{2}}{\partial z_{1} \partial z_{2}} + 2(h_{1}-h_{2})(z_{1}-z_{2})(\partial_{z_{1}}+\partial_{z_{2}}) +cc]$ which gives the mass term $C_{k}$ as the eigenvalue. Apart from this, the integral expression of (\ref{expression of B}) already assures the boundary condition for OPE block. The powerful feature of (\ref{4}) is the fact, that, it contains the well-known HKLL kernel of AdS$_{2}$ scalar field. It would be nice to see how one could get the connection of conserved current block with AdS$_{2}$ scalar field from the expression of (\ref{expression of B}).

\section{Explicit proof of intertwinement for geodesic operator/OPE block correspondence with unequal spin}\label{intert}
Using the intertwinement property of X-ray transform of a bulk field $\phi$, the authors of \cite{Czech:2016xec} have shown that the k-space equation of motion of scalar OPE block(for the OPE of two scalar primaries $\Delta_{i}=\Delta_{j}$) intertwines with the free bulk scalar Klein-Gordon equation.
\begin{align}\label{int}
2(\Box_{dS_{2}}+\Box_{\bar{dS_{2}}})R\phi = -R\Box_{AdS_{3}}\phi
\end{align}
Where $R \equiv \int_{\gamma}ds$ and $ds$ is the infinitesimal length of the geodesic $\gamma$. The main result of \cite{Czech:2016xec} is $B_{k} = R\phi$ which is argued from this intertwinement property by showing that they both obey the same EoM and the same boundary condition at $x_{1} \rightarrow x_{2}$ limit. On the other hand, for $\Delta_{i} \neq \Delta_{j}$, \cite{daCunha:2016crm} has argued that the form of OPE block becomes 
\begin{align}
B_{k}(x_{1},x_{2},\Delta_{i},\Delta_{j}) = \int_{\gamma}e^{-s\Delta_{ij}}\phi ds = R(e^{-s\Delta_{ij}}\phi)
\end{align}
This modified geodesic operator $R(e^{-s\Delta_{ij}}\phi)$ also satisfies the intertwinement property. As $s$ being a geodesic length, which is manifestly diffeomorphism invariant quantity, the isometry group element of AdS$_{3}$ does not change it. Thus it is straightforward to prove intertwinement of differential operator in k-space and AdS, i.e 
\begin{align}
\mathcal{L}_{k-space}R(e^{-s\Delta_{ij}}\phi) = -R(e^{-s\Delta_{ij}}\mathcal{L}_{AdS}\phi)
\end{align}
We will now see how this intertwinement property of Radon transform in CFT$_{2}$ holds, by explicitly working on Poincare AdS$_{3}$ on a constant time slice. More precisely, we want to show the following,
\begin{align}
[2(x_{1}-x_{2})^{2}\frac{\partial^{2}}{\partial x_{1} \partial x_{2}} + 2\Delta_{ij}(x_{1}-x_{2})(\partial_{x_{1}}+\partial_{x_{2}})]\int_{\gamma}e^{-s\Delta_{ij}}\phi ds = -\int_{\gamma}e^{-s\Delta_{ij}}\Box_{AdS_{2}}\phi ds
\end{align}
At the constant time slice there is no other constraint equation(e.g John's equation). We will use this result for arbitrary spin cases.\\
Let us now proceed by choosing Poincare coordinates. We define the geodesic length from the centre of the geodesic to some arbitrary point(x,z) on the geodesic. Therefore the geodesic distance is given by,
\begin{align}
s=\int_{\lambda_{1}}^{\lambda_{2}}\frac{\sqrt{\dot{x}^{2}+\dot{z}^{2}}}{z}d\lambda = \int_{\frac{\pi}{2}}^{\theta}d\theta \frac{1}{\sin\theta} = -\ln(\csc\theta +\cot\theta) \quad \textrm{and} \quad ds = \csc\theta d\theta
\end{align} 
where $x=\frac{(x_{1}-x_{2})}{2}\cos\theta+\frac{(x_{1}+x_{2})}{2}$ and $z=\frac{(x_{1}-x_{2})}{2}\sin\theta$. $x_{1},x_{2}$ are the end points of the geodesic. For technical simplicity we will use $r=\frac{\Delta}{2}=\frac{(x_{1}-x_{2})}{2}$ and $t=\frac{T}{2}=\frac{(x_{1}+x_{2})}{2}$\\
Let us first consider the action of $(x_{1}-x_{2})^{2}\frac{\partial^{2}}{\partial x_{1} \partial x_{2}}$ on geodesic operator. Since $s$ is only a function of $\theta$, $\frac{\partial^{2}}{\partial x_{1} \partial x_{2}}$ does not act on $s$.
\begin{align}\label{box}
&(x_{1}-x_{2})^{2}\frac{\partial^{2}}{\partial x_{1} \partial x_{2}}\int_{\gamma}ds e^{-s\Delta_{ij}}\phi  = \int_{\gamma}ds e^{-s\Delta_{ij}}4r^{2}[\frac{\partial^{2}}{\partial x_{1} \partial x_{2}}]\phi(r\sin\theta,r\cos\theta+t) \nonumber \\
&=  \int_{\gamma}ds e^{-s\Delta_{ij}}(4r^{2}\frac{\sin\theta}{2}[-\frac{\sin\theta}{2}\partial_{z}+\frac{1-\cos\theta}{2}\partial_{x}]\partial_{z}\phi(r\sin\theta,r\cos\theta+t) + \nonumber \\
&+ 4r^{2}\frac{1+\cos\theta}{2}[-\frac{\sin\theta}{2}\partial_{z}+\frac{1-\cos\theta}{2}\partial_{x}]\partial_{x}\phi(r\sin\theta,r\cos\theta+t) )\nonumber \\
&= \int_{\gamma}ds e^{-s\Delta_{ij}}\left(r^{2}\sin^{2}\theta\partial^{2}_{x} - r^{2}\sin^{2}\theta\partial^{2}_{z} - 2r^{2}\sin\theta\cos\theta\partial_{x}\partial_{z}\right)\phi(r\sin\theta,r\cos\theta+t)
\end{align}
We want to express $\partial_{x}\phi$ and $\partial_{z}\phi$ in terms of $\partial_{r}\phi$ and $\partial_{\theta}\phi$. Therefore, $\partial_{x}\phi = \cos\theta\partial_{r}\phi-\frac{\sin\theta}{r}\partial_{\theta}\phi$ and $\partial_{z}\phi = \sin\theta\partial_{r}\phi+\frac{\cos\theta}{r}\partial_{\theta}\phi$. Similarly we get,
\begin{align}
&r^{2}\sin^{2}\theta\partial^{2}_{x}\phi = [r^{2}\sin^{2}\theta\cos^{2}\theta\partial^{2}_{r} - 2r\sin^{3}\theta\cos\theta\partial_{r}\partial_{\theta} + \nonumber \\ &2\sin^{3}\theta\cos\theta\partial_{\theta} + r\sin^{4}\theta\partial_{r} + \sin^{4}\theta\partial^{2}_{\theta}]\phi \quad \textrm{and},  \\
&2r^{2}\sin\theta\cos\theta\partial_{x}\partial_{z}\phi = [2\sin^{2}\theta\cos^{2}\theta r^{2}\partial^{2}_{r} + 2\sin\theta\cos\theta r(1-2\sin^{2}\theta)\partial_{r}\partial_{\theta} - \nonumber \\
 &2\sin\theta\cos\theta(1-2\sin^{2}\theta)\partial_{\theta} - 2\sin^{2}\theta\cos^{2}\theta r\partial_{r} -2\sin^{2}\theta\cos^{2}\theta\partial^{2}_{\theta}]\phi 
\end{align}
Using the above expression, we find,
\begin{align}
&(r^{2}\sin^{2}\theta\partial^{2}_{x} - 2r^{2}\sin\theta\cos\theta\partial_{x}\partial_{z})\phi \nonumber \\
&= [-r^{2}\sin^{2}\theta\partial^{2}_{x} - 2\sin\theta\cos\theta r\partial_{r}\partial_{\theta} + 2\sin\theta\cos\theta\partial_{\theta} + 2\sin^{2}\theta r\partial_{r} + 2\sin^{2}\theta\partial^{2}_{\theta}]\phi 
\end{align}
Hence (\ref{box}) can be re-expressed as,
\begin{align}\label{first}
&(x_{1}-x_{2})^{2}\frac{\partial^{2}}{\partial x_{1} \partial x_{2}}\int_{\gamma}ds e^{-s\Delta_{ij}}\phi  = \int_{\gamma}ds e^{-s\Delta_{ij}}[-z^{2}\partial^{2}_{x} - z^{2}\partial^{2}_{z}]\phi - \nonumber \\
&\int_{\gamma}ds e^{-s\Delta_{ij}}[ 2\sin\theta\cos\theta r\partial_{r}\partial_{\theta} + 
  2\sin\theta\cos\theta\partial_{\theta} + 2\sin^{2}\theta r\partial_{r} + 2\sin^{2}\theta\partial^{2}_{\theta}]\phi \nonumber \\
 & = \int_{\gamma}ds e^{-s\Delta_{ij}}\Box_{AdS_{2}}\phi - 2\int_{\gamma}d\theta e^{-s\Delta_{ij}}\partial_{\theta}(\cos\theta r\partial_{r}\phi) - 2\int_{\gamma}d\theta e^{-s\Delta_{ij}}\sin\theta r\partial_{r}\phi + \nonumber \\
 & 2\int_{\gamma}d\theta e^{-s\Delta_{ij}}\partial_{\theta}(\cos\theta\phi) + \int_{\gamma}d\theta e^{-s\Delta_{ij}}\sin\theta\phi +2\int_{\gamma}d\theta e^{-s\Delta_{ij}}\sin\theta r\partial_{r}\phi + \nonumber \\
&  2\int_{\gamma}d\theta e^{-s\Delta_{ij}}\partial_{\theta}(\sin\theta\partial_{\theta}\phi) - 2\int_{\gamma}d\theta e^{-s\Delta_{ij}}\partial_{\theta}(\cos\theta\phi) - 2\int_{\gamma}d\theta e^{-s\Delta_{ij}}\sin\theta\phi \nonumber \\
&=\int_{\gamma}ds e^{-s\Delta_{ij}}\Box_{AdS_{2}}\phi-2\Delta_{ij}\int_{\gamma}d\theta e^{-s\Delta_{ij}}\cot\theta r\partial_{r}\phi + 2\Delta_{ij}\int_{\gamma}d\theta e^{-s\Delta_{ij}}\partial_{\theta}\phi + \textrm{(total derivative terms)}
\end{align}
Where $\Box_{AdS_{2}} =-z^{2}\partial^{2}_{x} - z^{2}\partial^{2}_{z}$.\\
Let us now consider the action of the other term $\Delta_{ij}(x_{1}-x_{2})(\partial_{x_{1}}+\partial_{x_{2}})$ on $\int_{\gamma}e^{-s\Delta_{ij}}\phi ds$. This gives,
\begin{align}\label{second}
&2\Delta_{ij}(x_{1}-x_{2})\partial_{T}\int_{\gamma}ds e^{-s\Delta_{ij}}\phi(z,x+\frac{T}{2}) \nonumber \\
&= \Delta_{ij}(x_{1}-x_{2})\int_{\gamma}ds e^{-s\Delta_{ij}}\partial_{x}\phi(z,x+\frac{T}{2})|_{z} \nonumber \\
&=\Delta_{ij}(x_{1}-x_{2})\int_{\gamma}ds e^{-s\Delta_{ij}} \left(-\frac{\sin\theta}{\frac{(x_{1}-x_{2})}{2}}\partial_{\theta}\phi + \cos\theta\partial_{r}\phi\right) \nonumber \\
&=-2\Delta_{ij}\int_{\gamma}d\theta e^{-s\Delta_{ij}}\partial_{\theta}\phi +2\Delta_{ij}\int_{\gamma}d\theta e^{-s\Delta_{ij}} \cot\theta r\partial_{r}\phi
\end{align}
Therefore, combining (\ref{first}) and (\ref{second}), we finally get the desired result of intertwining property,
\begin{align}\label{geodesic integral}
2[(x_{1}-x_{2})^{2}\frac{\partial^{2}}{\partial x_{1} \partial x_{2}} + \Delta_{ij}(x_{1}-x_{2})(\partial_{x_{1}}+\partial_{x_{2}})]\int_{\gamma}ds e^{-s\Delta_{ij}}\phi = -\int_{\gamma}ds e^{-s\Delta_{ij}}\Box_{AdS_{2}}\phi
\end{align}
For the conserved current case, the k-space equation in terms of modified geodesic operator we found, is the following,
\begin{align}
&[2(z_{1}-z_{2})^{2}\frac{\partial^{2}}{\partial z_{1} \partial z_{2}} + 2(\bar{z}_{1}-\bar{z}_{2})^{2}\frac{\partial^{2}}{\partial \bar{z}_{1} \partial \bar{z}_{2}} + 
2 h_{ij}(z_{1}-z_{2})(\partial_{z_{1}}+\partial_{z_{2}})+ 2 \bar{h}_{ij}(\bar{z}_{1}-\bar{z}_{2})(\partial_{\bar{z}_{1}}+\partial_{\bar{z}_{2}})]\nonumber \\
& \int_{\gamma}ds e^{-sh_{ij}}\phi_{AdS_{2}}(x(s)) \int_{\gamma'}ds' e^{-s'\bar{h}_{ij}}\phi_{AdS_{2}}(x'(s')) =C_{k}\int_{\gamma}ds e^{-sh_{ij}}\phi_{AdS_{2}}(x(s)) \int_{\gamma'}ds' e^{-s'\bar{h}_{ij}}\phi_{AdS_{2}}(x'(s'))
\end{align}
Here the geodesic operator decouples into two geodesic integrals of free scalar fields in AdS$_{2}$. Each holomorphic and anti-holomorphic part of k-space Laplacian acts on the geodesic integral similarly as (\ref{geodesic integral}). Using it we can find similarly,
\begin{align}
&[2(z_{1}-z_{2})^{2}\frac{\partial^{2}}{\partial z_{1} \partial z_{2}} + 2(\bar{z}_{1}-\bar{z}_{2})^{2}\frac{\partial^{2}}{\partial \bar{z}_{1} \partial \bar{z}_{2}} + 
2 h_{ij}(z_{1}-z_{2})(\partial_{z_{1}}+\partial_{z_{2}})+ \nonumber \\
&2 \bar{h}_{ij}(\bar{z}_{1}-\bar{z}_{2})(\partial_{\bar{z}_{1}}+\partial_{\bar{z}_{2}})] \int_{\gamma}ds e^{-sh_{ij}}\phi_{AdS_{2}}(x(s)) \int_{\gamma'}ds' e^{-s'\bar{h}_{ij}}\phi_{AdS_{2}}(x'(s'))  \nonumber \\
&= -\int_{\gamma}ds e^{-sh_{ij}}\int_{\gamma'}ds' e^{-s'\bar{h}_{ij}}[\phi'(x(s'))\Box_{AdS_{2}}\phi(x(s))+\phi(x(s))\Box_{\bar{AdS_{2}}}\phi'(x(s'))]
\end{align}
Therefore intertwinement of Laplacian also holds for the spinning geodesic operators. Next we want to check,  how the initial condition of field in k-space agrees with the AdS/CFT boundary condition via such geodesic operator, in this case.
\begin{align}
&\lim_{z_{1},\bar{z}_{1}\rightarrow z_{2},\bar{z}_{2}}B_{h_{k}}^{ij}(z_{1},z_{2};\bar{z}_{1},\bar{z}_{2})=(z_{1}-z_{2})^{h_{k}}(\bar{z}_{1}-\bar{z}_{2})^{\bar{h}_{k}}\mathcal{O}_{h_{k}}(z_{1})\mathcal{O}_{\bar{h}_{k}}(\bar{z}_{1}) \nonumber \\
&\sim \lim_{z\rightarrow 0}\phi_{AdS_{2}}(z_{1},z=z_{1}-z_{2})\lim_{z'\rightarrow 0}\phi_{AdS_{2}}(\bar{z}_{1},z'= \bar{z}_{1}-\bar{z}_{2}) \nonumber \\
&= \lim_{z_{1},\bar{z}_{1}\rightarrow z_{2},\bar{z}_{2}}R\phi_{AdS_{2}}(z(s),z_{1}(s))R\phi_{AdS_{2}}(z'(s'),\bar{z}_{1}(s'))
\end{align}
Hence, our result for expression of conserved current OPE block is consistent with the k-space approach of matching equation of motion and initial condition from both side in AdS/CFT.



\begin{thebibliography}{99}
 
  

  
  
  
  \bibitem{Dirac:1936fq} 
  P.~A.~M.~Dirac,
  ``Wave equations in conformal space,''
  Annals Math.\  {\bf 37}, 429 (1936).
  doi:10.2307/1968455
  
  \bibitem{Boulware:1970ty} 
  D.~G.~Boulware, L.~S.~Brown and R.~D.~Peccei,
  ``Deep-inelastic electroproduction and conformal symmetry,''
  Phys.\ Rev.\ D {\bf 2}, 293 (1970).
  doi:10.1103/PhysRevD.2.293
  
  \bibitem{Cornalba:2009ax} 
  L.~Cornalba, M.~S.~Costa and J.~Penedones,
  ``Deep Inelastic Scattering in Conformal QCD,''
  JHEP {\bf 1003}, 133 (2010)
  doi:10.1007/JHEP03(2010)133
  [arXiv:0911.0043 [hep-th]].
  
  \bibitem{Weinberg:2010fx} 
  S.~Weinberg,
  ``Six-dimensional Methods for Four-dimensional Conformal Field Theories,''
  Phys.\ Rev.\ D {\bf 82}, 045031 (2010)
  doi:10.1103/PhysRevD.82.045031
  [arXiv:1006.3480 [hep-th]].
  
  \bibitem{Rychkov:2016iqz} 
  S.~Rychkov,
  ``EPFL Lectures on Conformal Field Theory in D>= 3 Dimensions,''
  doi:10.1007/978-3-319-43626-5
  arXiv:1601.05000 [hep-th].
  
  \bibitem{Penedones:2016voo} 
  J.~Penedones,
  ``TASI lectures on AdS/CFT,''
  [arXiv:1608.04948 [hep-th]].
  
  \bibitem{Hijano:2015zsa} 
  E.~Hijano, P.~Kraus, E.~Perlmutter and R.~Snively,
  ``Witten Diagrams Revisited: The AdS Geometry of Conformal Blocks,''
  JHEP {\bf 1601}, 146 (2016)
  doi:10.1007/JHEP01(2016)146
  [arXiv:1508.00501 [hep-th]].
  
\bibitem{Mack:1969rr} 
  G.~Mack and A.~Salam,
  ``Finite component field representations of the conformal group,''
  Annals Phys.\  {\bf 53}, 174 (1969).
  doi:10.1016/0003-4916(69)90278-4
  
  \bibitem{Ferrara:1973eg} 
  S.~Ferrara, R.~Gatto and A.~F.~Grillo,
  ``Conformal algebra in space-time and operator product expansion,''
  Springer Tracts Mod.\ Phys.\  {\bf 67}, 1 (1973).
  doi:10.1007/BFb0111104
  
  \bibitem{Ferrara:1973yt} 
  S.~Ferrara, A.~F.~Grillo and R.~Gatto,
  ``Tensor representations of conformal algebra and conformally covariant operator product expansion,''
  Annals Phys.\  {\bf 76}, 161 (1973).
  doi:10.1016/0003-4916(73)90446-6  
  
  \bibitem{Costa:2011mg} 
  M.~S.~Costa, J.~Penedones, D.~Poland and S.~Rychkov,
  ``Spinning Conformal Correlators,''
  JHEP {\bf 1111}, 071 (2011)
  doi:10.1007/JHEP11(2011)071
  [arXiv:1107.3554 [hep-th]].
  
  \bibitem{Costa:2011dw} 
  M.~S.~Costa, J.~Penedones, D.~Poland and S.~Rychkov,
  ``Spinning Conformal Blocks,''
  JHEP {\bf 1111}, 154 (2011)
  doi:10.1007/JHEP11(2011)154
  [arXiv:1109.6321 [hep-th]].
  
  \bibitem{SimmonsDuffin:2012uy} 
  D.~Simmons-Duffin,
  ``Projectors, Shadows, and Conformal Blocks,''
  JHEP {\bf 1404}, 146 (2014)
  doi:10.1007/JHEP04(2014)146
  [arXiv:1204.3894 [hep-th]].
  
  \bibitem{Alday:2013cwa} 
  L.~F.~Alday and A.~Bissi,
  ``Higher-spin correlators,''
  JHEP {\bf 1310}, 202 (2013)
  doi:10.1007/JHEP10(2013)202
  [arXiv:1305.4604 [hep-th]].
  
  \bibitem{Costa:2014rya} 
  M.~S.~Costa and T.~Hansen,
  ``Conformal correlators of mixed-symmetry tensors,''
  JHEP {\bf 1502}, 151 (2015)
  doi:10.1007/JHEP02(2015)151
  [arXiv:1411.7351 [hep-th]].
  
  \bibitem{Penedones:2015aga} 
  J.~Penedones, E.~Trevisani and M.~Yamazaki,
  ``Recursion Relations for Conformal Blocks,''
  JHEP {\bf 1609}, 070 (2016)
  doi:10.1007/JHEP09(2016)070
  [arXiv:1509.00428 [hep-th]].
  
  \bibitem{Rejon-Barrera:2015bpa} 
  F.~Rejon-Barrera and D.~Robbins,
  ``Scalar-Vector Bootstrap,''
  JHEP {\bf 1601}, 139 (2016)
  doi:10.1007/JHEP01(2016)139
  [arXiv:1508.02676 [hep-th]].
  
  \bibitem{Iliesiu:2015qra} 
  L.~Iliesiu, F.~Kos, D.~Poland, S.~S.~Pufu, D.~Simmons-Duffin and R.~Yacoby,
  ``Bootstrapping 3D Fermions,''
  JHEP {\bf 1603}, 120 (2016)
  doi:10.1007/JHEP03(2016)120
  [arXiv:1508.00012 [hep-th]].
  
  \bibitem{Iliesiu:2015akf} 
  L.~Iliesiu, F.~Kos, D.~Poland, S.~S.~Pufu, D.~Simmons-Duffin and R.~Yacoby,
  ``Fermion-Scalar Conformal Blocks,''
  JHEP {\bf 1604}, 074 (2016)
  doi:10.1007/JHEP04(2016)074
  [arXiv:1511.01497 [hep-th]].
  
  \bibitem{Echeverri:2015rwa} 
  A.~Castedo Echeverri, E.~Elkhidir, D.~Karateev and M.~Serone,
  ``Deconstructing Conformal Blocks in 4D CFT,''
  JHEP {\bf 1508}, 101 (2015)
  doi:10.1007/JHEP08(2015)101
  [arXiv:1505.03750 [hep-th]].
  
  \bibitem{Costa:2016hju} 
  M.~S.~Costa, T.~Hansen, J.~Penedones and E.~Trevisani,
  ``Projectors and seed conformal blocks for traceless mixed-symmetry tensors,''
  JHEP {\bf 1607}, 018 (2016)
  doi:10.1007/JHEP07(2016)018
  [arXiv:1603.05551 [hep-th]].
  
  \bibitem{Costa:2016xah} 
  M.~S.~Costa, T.~Hansen, J.~Penedones and E.~Trevisani,
  ``Radial expansion for spinning conformal blocks,''
  JHEP {\bf 1607}, 057 (2016)
  doi:10.1007/JHEP07(2016)057
  [arXiv:1603.05552 [hep-th]].
  
  \bibitem{Schomerus:2016epl} 
  V.~Schomerus, E.~Sobko and M.~Isachenkov,
  ``Harmony of Spinning Conformal Blocks,''
  JHEP {\bf 1703}, 085 (2017)
  doi:10.1007/JHEP03(2017)085
  [arXiv:1612.02479 [hep-th]].
  
  \bibitem{Karateev:2017jgd} 
  D.~Karateev, P.~Kravchuk and D.~Simmons-Duffin,
  ``Weight Shifting Operators and Conformal Blocks,''
  JHEP {\bf 1802}, 081 (2018)
  doi:10.1007/JHEP02(2018)081
  [arXiv:1706.07813 [hep-th]].

  
  \bibitem{Dyer:2017zef} 
  E.~Dyer, D.~Z.~Freedman and J.~Sully,
  ``Spinning Geodesic Witten Diagrams,''
  arXiv:1702.06139 [hep-th].
  
  \bibitem{Castro:2017hpx} 
  A.~Castro, E.~Llabrés and F.~Rejon-Barrera,
  ``Geodesic Diagrams, Gravitational Interactions \& OPE Structures,''
  JHEP {\bf 1706}, 099 (2017)
  doi:10.1007/JHEP06(2017)099
  [arXiv:1702.06128 [hep-th]].
  
  \bibitem{Nishida:2016vds} 
  M.~Nishida and K.~Tamaoka,
  ``Geodesic Witten diagrams with an external spinning field,''
  PTEP {\bf 2017}, no. 5, 053B06 (2017)
  doi:10.1093/ptep/ptx055
  [arXiv:1609.04563 [hep-th]].
  
  \bibitem{Sleight:2017fpc} 
  C.~Sleight and M.~Taronna,
  ``Spinning Witten Diagrams,''
  JHEP {\bf 1706}, 100 (2017)
  doi:10.1007/JHEP06(2017)100
  [arXiv:1702.08619 [hep-th]].
  
  \bibitem{Chen:2017yia} 
  H.~Y.~Chen, E.~J.~Kuo and H.~Kyono,
  ``Anatomy of Geodesic Witten Diagrams,''
  JHEP {\bf 1705}, 070 (2017)
  doi:10.1007/JHEP05(2017)070
  [arXiv:1702.08818 [hep-th]].
  
  \bibitem{Tamaoka:2017jce} 
  K.~Tamaoka,
  ``Geodesic Witten diagrams with anti-symmetric exchange,''
  Phys.\ Rev.\ D {\bf 96}, 086007 (2017)
  doi:10.1103/PhysRevD.96.086007
  [arXiv:1707.07934 [hep-th]].
  
  \bibitem{Nishida:2018opl} 
  M.~Nishida and K.~Tamaoka,
  ``Fermions in Geodesic Witten Diagrams,''
  JHEP {\bf 1807}, 149 (2018)
  doi:10.1007/JHEP07(2018)149
  [arXiv:1805.00217 [hep-th]].
  
  \bibitem{Costa:2018mcg} 
  M.~S.~Costa and T.~Hansen,
  ``AdS Weight Shifting Operators,''
  JHEP {\bf 1809}, 040 (2018)
  doi:10.1007/JHEP09(2018)040
  [arXiv:1805.01492 [hep-th]].

  
  \bibitem{Bhatta:2016hpz} 
  A.~Bhatta, P.~Raman and N.~V.~Suryanarayana,
  ``Holographic Conformal Partial Waves as Gravitational Open Wilson Networks,''
  JHEP {\bf 1606}, 119 (2016)
  doi:10.1007/JHEP06(2016)119
  [arXiv:1602.02962 [hep-th]].
  
  
  
  \bibitem{Besken:2016ooo} 
  M.~Besken, A.~Hegde, E.~Hijano and P.~Kraus,
  ``Holographic conformal blocks from interacting Wilson lines,''
  JHEP {\bf 1608}, 099 (2016)
  doi:10.1007/JHEP08(2016)099
  [arXiv:1603.07317 [hep-th]].
  
  \bibitem{Bhatta:2018gjb} 
  A.~Bhatta, P.~Raman and N.~V.~Suryanarayana,
  ``Scalar Blocks as Gravitational Wilson Networks,''
  arXiv:1806.05475 [hep-th].
  
\bibitem{Czech:2016xec} 
  B.~Czech, L.~Lamprou, S.~McCandlish, B.~Mosk and J.~Sully,
  ``A Stereoscopic Look into the Bulk,''
  JHEP {\bf 1607}, 129 (2016)
  doi:10.1007/JHEP07(2016)129
  [arXiv:1604.03110 [hep-th]].
  
\bibitem{deBoer:2016pqk} 
  J.~de Boer, F.~M.~Haehl, M.~P.~Heller and R.~C.~Myers,
  ``Entanglement, holography and causal diamonds,''
  JHEP {\bf 1608}, 162 (2016)
  doi:10.1007/JHEP08(2016)162
  [arXiv:1606.03307 [hep-th]].
  
\bibitem{Czech:2015qta} 
  B.~Czech, L.~Lamprou, S.~McCandlish and J.~Sully,
  ``Integral Geometry and Holography,''
  JHEP {\bf 1510}, 175 (2015)
  doi:10.1007/JHEP10(2015)175
  [arXiv:1505.05515 [hep-th]].
 
  
\bibitem{daCunha:2016crm}
B.~Carneiro da Cunha and M.~Guica,
`` Exploring the BTZ bulk with boundary conformal blocks,''
[arXiv:1604.07383 [hep-th]].

\bibitem{Karch:2017fuh} 
  A.~Karch, J.~Sully, C.~F.~Uhlemann and D.~G.~E.~Walker,
  ``Boundary Kinematic Space,''
  JHEP {\bf 1708}, 039 (2017)
  doi:10.1007/JHEP08(2017)039
  [arXiv:1703.02990 [hep-th]].
  
  \bibitem{Cresswell:2017mbk} 
  J.~C.~Cresswell and A.~W.~Peet,
  ``Kinematic space for conical defects,''
  JHEP {\bf 1711}, 155 (2017)
  doi:10.1007/JHEP11(2017)155
  [arXiv:1708.09838 [hep-th]].
  
  \bibitem{Fukuda:2017cup} 
  M.~Fukuda, N.~Kobayashi and T.~Nishioka,
  ``Operator product expansion for conformal defects,''
  JHEP {\bf 1801}, 013 (2018)
  doi:10.1007/JHEP01(2018)013
  [arXiv:1710.11165 [hep-th]].
  
  \bibitem{Cresswell:2018mpj} 
  J.~C.~Cresswell, I.~T.~Jardine and A.~W.~Peet,
  ``Holographic relations for OPE blocks in excited states,''
  arXiv:1809.09107 [hep-th].
  
  \bibitem{Dolan:2003hv} 
  F.~A.~Dolan and H.~Osborn,
  ``Conformal partial waves and the operator product expansion,''
  Nucl.\ Phys.\ B {\bf 678}, 491 (2004)
  doi:10.1016/j.nuclphysb.2003.11.016
  [hep-th/0309180].
  
  \bibitem{Dolan:2000ut} 
  F.~A.~Dolan and H.~Osborn,
  ``Conformal four point functions and the operator product expansion,''
  Nucl.\ Phys.\ B {\bf 599}, 459 (2001)
  doi:10.1016/S0550-3213(01)00013-X
  [hep-th/0011040].
  
   \bibitem{Osborn:2012vt} 
  H.~Osborn,
  ``Conformal Blocks for Arbitrary Spins in Two Dimensions,''
  Phys.\ Lett.\ B {\bf 718}, 169 (2012)
  doi:10.1016/j.physletb.2012.09.045
  [arXiv:1205.1941 [hep-th]].
  
\bibitem{Dolan:2011dv} 
  F.~A.~Dolan and H.~Osborn,
  ``Conformal Partial Waves: Further Mathematical Results,''
  arXiv:1108.6194 [hep-th].  
  
\bibitem{Ferrara:1971vh} 
  S.~Ferrara, A.~F.~Grillo and R.~Gatto,
 ``Manifestly conformal covariant operator-product expansion,''
  Lett.\ Nuovo Cim.\  {\bf 2S2}, 1363 (1971)
  [Lett.\ Nuovo Cim.\  {\bf 2}, 1363 (1971)].
  doi:10.1007/BF02770435
  
\bibitem{Ferrara:1973vz} 
  S.~Ferrara, A.~F.~Grillo, G.~Parisi and R.~Gatto,
  ``Covariant expansion of the conformal four-point function,''
  Nucl.\ Phys.\ B {\bf 49}, 77 (1972)
  Erratum: [Nucl.\ Phys.\ B {\bf 53}, 643 (1973)].
  doi:10.1016/0550-3213(72)90587-1, 10.1016/0550-3213(73)90467-7
  
  
  \bibitem{Zamolodchikov:1985ie} 
  A.~B.~Zamolodchikov,
  ``Conformal Symmetry In Two-dimensions: An Explicit Recurrence Formula For The Conformal Partial Wave Amplitude,''
  Commun.\ Math.\ Phys.\  {\bf 96}, 419 (1984).
  doi:10.1007/BF01214585.
  
  \bibitem{Fitzpatrick:2014vua} 
  A.~L.~Fitzpatrick, J.~Kaplan and M.~T.~Walters,
  ``Universality of Long-Distance AdS Physics from the CFT Bootstrap,''
  JHEP {\bf 1408}, 145 (2014)
  doi:10.1007/JHEP08(2014)145
  [arXiv:1403.6829 [hep-th]].
  
  \bibitem{Hijano:2015rla} 
  E.~Hijano, P.~Kraus and R.~Snively,
  ``Worldline approach to semi-classical conformal blocks,''
  JHEP {\bf 1507}, 131 (2015)
  doi:10.1007/JHEP07(2015)131
  [arXiv:1501.02260 [hep-th]].
  
  \bibitem{Alkalaev:2015wia} 
  K.~B.~Alkalaev and V.~A.~Belavin,
  ``Classical conformal blocks via AdS/CFT correspondence,''
  JHEP {\bf 1508}, 049 (2015)
  doi:10.1007/JHEP08(2015)049
  [arXiv:1504.05943 [hep-th]].
  
  \bibitem{Hijano:2015qja} 
  E.~Hijano, P.~Kraus, E.~Perlmutter and R.~Snively,
  ``Semiclassical Virasoro blocks from AdS$_{3}$ gravity,''
  JHEP {\bf 1512}, 077 (2015)
  doi:10.1007/JHEP12(2015)077
  [arXiv:1508.04987 [hep-th]].
  
  \bibitem{Hamilton:2005ju} 
  A.~Hamilton, D.~N.~Kabat, G.~Lifschytz and D.~A.~Lowe,
  ``Local bulk operators in AdS/CFT: A Boundary view of horizons and locality,''
  Phys.\ Rev.\ D {\bf 73}, 086003 (2006)
  doi:10.1103/PhysRevD.73.086003
  [hep-th/0506118].
  
  \bibitem{Hamilton:2006az} 
  A.~Hamilton, D.~N.~Kabat, G.~Lifschytz and D.~A.~Lowe,
  ``Holographic representation of local bulk operators,''
  Phys.\ Rev.\ D {\bf 74}, 066009 (2006)
  doi:10.1103/PhysRevD.74.066009
  [hep-th/0606141].  

  
  
  \bibitem{Sarkar:2014dma} 
  D.~Sarkar and X.~Xiao,
  ``Holographic Representation of Higher Spin Gauge Fields,''
  Phys.\ Rev.\ D {\bf 91}, no. 8, 086004 (2015)
  doi:10.1103/PhysRevD.91.086004
  [arXiv:1411.4657 [hep-th]].
 
  
  \bibitem{Czech:2016tqr} 
  B.~Czech, L.~Lamprou, S.~McCandlish, B.~Mosk and J.~Sully,
  ``Equivalent Equations of Motion for Gravity and Entropy,''
  JHEP {\bf 1702}, 004 (2017)
  doi:10.1007/JHEP02(2017)004
  [arXiv:1608.06282 [hep-th]].
  
  
  
  \bibitem{Kabat:2012hp} 
  D.~Kabat, G.~Lifschytz, S.~Roy and D.~Sarkar,
  ``Holographic representation of bulk fields with spin in AdS/CFT,''
  Phys.\ Rev.\ D {\bf 86}, 026004 (2012)
  doi:10.1103/PhysRevD.86.026004, 10.1103/PhysRevD.86.029901
  [arXiv:1204.0126 [hep-th]].
  
  \bibitem{Heemskerk:2012np} 
  I.~Heemskerk,
  ``Construction of Bulk Fields with Gauge Redundancy,''
  JHEP {\bf 1209}, 106 (2012)
  doi:10.1007/JHEP09(2012)106
  [arXiv:1201.3666 [hep-th]].
  

  
  \bibitem{Vasiliev:1990en} 
  M.~A.~Vasiliev,
  ``Consistent equation for interacting gauge fields of all spins in (3+1)-dimensions,''
  Phys.\ Lett.\ B {\bf 243}, 378 (1990).
  doi:10.1016/0370-2693(90)91400-6
  
  \bibitem{Vasiliev:2003ev} 
  M.~A.~Vasiliev,
  ``Nonlinear equations for symmetric massless higher spin fields in (A)dS(d),''
  Phys.\ Lett.\ B {\bf 567}, 139 (2003)
  doi:10.1016/S0370-2693(03)00872-4
  
  \bibitem{Giombi:2016ejx} 
  S.~Giombi,
  ``Higher Spin — CFT Duality,''
  arXiv:1607.02967 [hep-th].
  
  \bibitem{Vasiliev:1995dn} 
  M.~A.~Vasiliev,
  ``Higher spin gauge theories in four-dimensions, three-dimensions, and two-dimensions,''
  Int.\ J.\ Mod.\ Phys.\ D {\bf 5}, 763 (1996)
  doi:10.1142/S0218271896000473
  [hep-th/9611024].
  
  \bibitem{Mikhailov:2002bp} 
  A.~Mikhailov,
  ``Notes on higher spin symmetries,''
  hep-th/0201019.
  
  
  \bibitem{Ferrara:1972xe} 
  S.~Ferrara and G.~Parisi,
  ``Conformal covariant correlation functions,''
  Nucl.\ Phys.\ B {\bf 42}, 281 (1972).
  doi:10.1016/0550-3213(72)90480-4
  
  \bibitem{Ferrara:1972ay} 
  S.~Ferrara, A.~F.~Grillo and G.~Parisi,
  ``Nonequivalence between conformal covariant wilson expansion in euclidean and minkowski space,''
  Lett.\ Nuovo Cim.\  {\bf 5S2}, 147 (1972)
  [Lett.\ Nuovo Cim.\  {\bf 5}, 147 (1972)].
  doi:10.1007/BF02815915
  
  \bibitem{Ferrara:1972uq} 
  S.~Ferrara, A.~F.~Grillo, G.~Parisi and R.~Gatto,
  ``The shadow operator formalism for conformal algebra. vacuum expectation values and operator products,''
  Lett.\ Nuovo Cim.\  {\bf 4S2}, 115 (1972)
  [Lett.\ Nuovo Cim.\  {\bf 4}, 115 (1972)].
  doi:10.1007/BF02907130
  
  
\end{thebibliography}
\end{document}